\documentclass[12pt, draftclsnofoot, onecolumn]{IEEEtran}
\usepackage{times}
\IEEEoverridecommandlockouts

\usepackage{cite}

 \usepackage{amsmath,amssymb,amsfonts,amsthm}
\usepackage{bbm}

 \usepackage{algorithmic, custom_macros}
 
 \usepackage{graphicx}
 \usepackage{mathtools}
\usepackage{multicol}
\usepackage[bookmarks=true]{hyperref}
\usepackage{xcolor}
\usepackage{tikz}
 \usepackage{tikz-3dplot}
 \usetikzlibrary{shapes.multipart}
 \usetikzlibrary{shapes.symbols}
\usepackage{pgfplots}

\usepackage{standalone}

\usepackage{subcaption}
\usepackage{comment}

\usepackage[
  plainruled,
  linesnumbered,
]{algorithm2e}

\SetAlgoCaptionLayout{centerline}
\DontPrintSemicolon

\makeatletter
\def\@algocf@capt@plainruled{above}
\renewcommand{\algocf@caption@plainruled}{%
  \vskip\AlCapSkip%
  \box\algocf@capbox%
  \vskip 5\algoheightrule}%
\makeatother

\usepackage{etoolbox}
\usepackage{float}
\usepackage[belowskip=-15pt,aboveskip=5pt]{caption}

\allowdisplaybreaks

\newcommand{\my}[1]{{\textcolor{black}{#1}}}

\newcommand\Alpha{\mathrm{A}}
\newtheorem{lemma}{Lemma}

\newtheorem{theorem}{Theorem}

\title{Robust Federated Learning with Connectivity Failures: A Semi-Decentralized Framework with Collaborative Relaying}

\author{Michal Yemini, Rajarshi Saha, Emre Ozfatura, \\Deniz G\"{u}nd\"{u}z and Andrea J. Goldsmith
\thanks{M.~Yemini and A.~J.~Goldsmith are  with the Faculty of 
Electrical and Computer Engineering, Princeton University. (Emails: myemini@princeton.edu, goldsmith@princeton.edu).
R.~Saha is with the Department of Electrical Engineering, Stanford University. (Email: rajsaha@stanford.edu).
E.~Ozfatura and D.~G\"{u}nd\"{u}z are with the Department of Electrical and Electronic Engineering, Imperial College London. (Emails: m.ozfatura@imperial.ac.uk, d.gunduz@imperial.ac.uk).}
\thanks{
M.~Yemini, R.~Saha, and A.~J.~Goldsmith are partially supported by the AFOSR award \#002484665 and a Huawei Intelligent Spectrum grant.
E.~Ozfatura and D.~G\"{u}nd\"{u}z received funding from the European Research Council (ERC) through Starting Grant BEACON (no. 677854) and the UK EPSRC (grant no. EP/T023600/1) under the CHIST-ERA program.}
\thanks{A preliminary version of this paper was presented  at the  2022 IEEE International Symposium on Information Theory (ISIT), \cite{our_ISIT_version}.}
}
\date{\today}

\begin{document}

\maketitle
\begin{abstract}

Intermittent connectivity of clients to the parameter server (PS) is a major bottleneck in federated edge learning frameworks.
The lack of constant connectivity induces a large generalization gap, especially when the local data distribution amongst clients exhibits heterogeneity.
To overcome intermittent communication outages between clients and the central PS, we introduce the concept of collaborative relaying wherein the participating clients relay their neighbors' local updates to the PS in order to boost the participation of clients with poor connectivity to the PS.
We propose a semi-decentralized federated learning framework in which at every communication round, each client initially computes a local consensus of a subset of its neighboring clients' updates, and eventually transmits to the PS a weighted average of its own update and those of its neighbors'.
We appropriately optimize these local consensus weights to ensure that the global update at the PS is unbiased with minimal variance -- consequently improving the convergence rate.
Numerical evaluations on the CIFAR-10 dataset demonstrate that our collaborative relaying approach outperforms federated averaging-based benchmarks for learning over intermittently-connected networks such as when the clients communicate over millimeter wave channels with intermittent blockages.
\end{abstract}

\section{Introduction}

Federated learning (FL) algorithms iteratively optimize a common objective function to learn a shared model over data samples that are localized over multiple distributed clients \cite{FL1}. 
FL approaches aim to reduce the required communication overhead and improve clients' privacy by letting each client train a local model on its local (private) dataset and forwarding them periodically to a centralized parameter server (PS).

In practical FL scenarios, some clients are stragglers and cannot send their estimates regularly, either because: $(i)$ they cannot finish their computation within a prescribed deadline, or $(ii)$ they cannot transmit their estimate to the PS successfully due to communication limitations \cite{chen2021distributed}.
Specifically, clients can suffer from intermittent connectivity to the PS, where their wireless communication channel is temporary blocked \cite{6834753,7511572,8047278,yasamin,pappas,zavlanos,gilIJRR}.
Stragglers deteriorate the convergence of FL as the computed local estimates become stale which reduced accuracy and can even result in bias in the final model in the case of persistent stragglers.
However, the case of communication stragglers that are limited due to loss of direct communication opportunities to the PS is inherently different from those that result from limited computation resources at the client, since the former can be solved by relaying the updates via neighboring clients.

Communication channel quality is considered a key design principle  in the federated edge learning (FEEL) framework \cite{oac0}, which takes into account the wireless  channel characteristics from the clients to the PS to optimize the convergence and final model performance at the PS. 
So far the FEEL paradigm has mainly focused on  direct  communication  from the clients to the PS, and aimed at improving the performance by resource allocation across clients  \cite{FL.CS-RA1,FL.CS-RA2,FL.CS-RA3,FL.CS-RA4,oac0,oac1,oac2,oac3,OTA_heterogenous_data_2022,CoBAAF_2022,oac_privacy1,oac_privacy2, FL.CS-RA7}; this model ignores possible cooperation among clients in the case of intermittent communication blockages.  \my{Robustness to channel blockages is critical for the reliable operations of mmWave communication systems and  robotic systems. While mmWave systems have many advantages in terms of utilizing a previously unused section of the radio spectrum for communication purposes, it is known to suffer from temporary channel blockage due to moving bodies \cite{6834753,7511572,8047278}. Additionally, overcoming channel blockage is  highly relevant for robotic networks where mobile robots explore remote parts of the area of interests and thus can be disconnected from a central server  \cite{yasamin,pappas,zavlanos,gilIJRR}. }

Relaying, also known as cooperative communications, has been widely studied in the literature in order to improve the coverage \cite{Kramer:now:06}, diversity \cite{Sendonaris:TCOM:03}, end-to-end signal quality \cite{Shutoy:JSTSP:07, Gunduz:TIT:13} as well as connectivity \cite{yemini_et_al:globeom2020,yemini_et_al:TWC_cloud_cluster} of wireless terminals. In FEEL, relaying has been considered previously in the context of hierarchical FL \cite{HFL2} to bring the clients closer to the PS with the aim of reducing the latency and interference. In this paper,  motivated by our prior works \cite{yemini_et_al:globeom2020,yemini_et_al:TWC_cloud_cluster}, where client cooperation is used to improve the connectivity to the cloud and to reduce the latency and scheduling overhead, we propose and analyze a new FEEL paradigm, where the clients cooperate to mitigate the detrimental effects of communication stragglers. 
In the proposed method, clients share with each other their current updates so that each client can send to the PS a weighted average of its own current update and those of its neighbors.
Thanks to relaying, the PS receives new updates from clients with intermittently failing  uplink connections, which would otherwise become stale and be discarded.
Moreover, we optimize the averaging weights in order to ensure that the transmitted updates to the PS $(i)$ achieve weak unbiasedness that preserves the objective function at the PS, and $(ii)$ minimize the convergence time of the learning algorithm.
We provide theoretical guarantees for the improvement in convergence rate of our scheme.

Most existing works on FL assume error-free rate-limited orthogonal communication links, assuming that the wireless imperfections are taken care of by an underlying communication protocol.
However, such a separation between the communication medium and the learning protocol can be strictly suboptimal \cite{oac0}.  
An alternative approach is to treat the communication of the model updates to the PS as an uplink communication problem and jointly optimize the learning algorithm and the communication scheme, taking wireless channel imperfections into account \cite{oac0,JSAC_Chen_2021}.
Within this framework, an original and promising approach is the  {\em over-the-air computation (OAC)} \cite{oac1,oac2,oac3,OTA_heterogenous_data_2022,CoBAAF_2022}, which exploits the signal superposition property of the wireless medium to convey the sum of the model updates in an uncoded fashion. 
In addition to bandwidth efficiency, the OAC framework also provides a certain level of anonymity to the clients due to its superposition nature; and hence, enhances the privacy of the participating clients \cite{oac_privacy1,oac_privacy2}.
We emphasize here that, in the OAC framework, the PS receives the aggregate model, and it is not possible to disentangle the individual model updates. Therefore, any strategy that utilizes a PS side aggregation mechanism with individual model updates to address unequal client participation is not compatible with the OAC framework.
One of the major advantages of our proposed scheme is that it mitigates the drawbacks of unequal client participation without requiring the identity of the transmitting clients or their individual updates at the PS.
Therefore, our solution is compatible with OAC, and can retain its bandwidth efficiency and anonymity benefits.

\paragraph*{Related works}

The conventional FL framework \cite{FL1} is orchestrated by a centralized entity called PS, which helps participating clients to reach a consensus on the model parameters by aggregating their locally trained models.
However, such a consensus mechanism requires the exchange of a large number of model parameters between the PS and clients, thereby   creating significant communication overhead. A decentralized learning framework has been introduced as an alternative to FL, in which the PS is removed to mitigate a potential communication bottleneck and a single point of failure.
In decentralized learning, each client shares its local model with the neighboring clients through device-to-device (D2D) communications, and model aggregation is executed at each client in parallel. In a sense, in decentralized learning, each client becomes a PS.
The aggregation strategy at each client is determined according to the network topology, that is the connection pattern between the clients, and often a fixed topology is considered \cite{decent1,decent2,decent3,decent4,decent5,decent6,decent7,decent8, decent9,Distributed_optimization_CDC_version}. 
These results can be further extended to  scenarios with time varying topologies \cite{decent_top1,decent_top2,decent_top3}. \\
\indent An alternative approach to both centralized and decentralized schemes is the {\em hierarchical FL (HFL)} framework \cite{HFL1,HFL2,HFL3,HFL4}, where multiple PSs are employed for the aggregation to prevent a communication bottleneck.
In HFL, clients are divided into clusters and a PS is assigned to each cluster to perform local aggregation, while the aggregated models at the clusters are later aggregated at the main PS in a subsequent step to obtain the global model. 
This framework has significant advantages over centralized and decentralized schemes,  particularly when the communication takes place over wireless channels since it allows spatial reuse of available resources \cite{HFL2}.\\
\indent Although HFL has certain advantages, this framework requires employing multiple PSs that may not be practical in certain scenarios.
Instead, the idea of hierarchical collaborative learning can be redesigned to combine hierarchical and decentralized  learning concepts, which is referred to as  {\em semi-decentralized FL}, where the local consensus follows decentralized learning with D2D communication, whereas the global consensus is orchestrated by the PS \cite{semi-decent1,semi-decent2}. One of the major challenges in FL that is not considered in the aforementioned works on semi-decentralized FL is the partial client connectivity \cite{part_device1,part_device2}. 
Unequal client participation due to intermittent connectivity exacerbates the impact of data heterogeneity \cite{FL.noniid1,FL.noniid2,FL.noniid3,FL.noniid4}, and increases the generalization gap.\\
\indent The connectivity of the clients is a particularly significant challenge in FEEL, where the clients and the PS communicate over unreliable wireless channels. 
Due to their diverse physical environments and distances to the PS, clients may have different connectivity to each other and the PS.
This problem has been recently addressed in \cite{FL.CS,FL.CS-RA1,FL.CS-RA2,FL.CS-RA3,FL.CS-RA4,FL.CS-RA5,FL.CS-RA6,FL.CS-RA7,Ozfatura_SPAWC_2021} by considering customized client selection mechanisms to seek a balance between the participation of the clients and the latency for the model aggregation in order to speed up the learning process. 
In this work, we adopt a different approach to the connectivity problem, and instead of designing a client selection mechanism, or optimizing resource allocation to balance client participation, we introduce a {\em relaying} mechanism that takes into account the nature of individual clients' connectivity to the PS and ensures that, in case of poor connectivity, their local information is conveyed to the PS with the help of their neighboring clients.

Finally, another related body of work considers coded computation as a possible solution to mitigate stragglers in distributed learning \cite{gradient_coding_ISIT_2016,gradient_coding_2017_ICML,gradient_coding_IT_2018,gradient_coding_entropy_2020, Amiri:TSP:19,Coded_FL_2021}. However, in the classical approach \cite{gradient_coding_ISIT_2016,gradient_coding_2017_ICML,gradient_coding_IT_2018,gradient_coding_entropy_2020, Amiri:TSP:19}, data is strategically allocated to clients to create redundant computations that can be exploited by the PS using additional information about the stragglers' identities.  The work \cite{Coded_FL_2021} incorporates coded computing into  FL  by sharing additional information about client's features with the PS, this can cause additional privacy leakage. Furthermore, the PS cannot be blind to the client's identities and thus this scheme cannot be implemented OAC straightforwardly.
Our approach does not require sharing additional client's information beside the client's updates. Finally, we note that conventional coded computation schemes spread the data among clients without constraints on the allowed combinations. In contrast, in our scheme the clients cannot choose their neighbors as they depend on the channel characteristics, thus coded computing techniques cannot be readily adopted.   

\subsection{Main Contributions} 
The main contributions of our paper can be summarized as follows:
\begin{itemize}
\item We propose a new semi-decentralized FL framework, which exploits local connections between clients to relay each other's estimates to the PS, to mitigate the negative impacts of intermittent client-PS connections on the learning performance. In the proposed framework. clients cooperate not only to learn a common model by exchanging model updates with the PS, but also to improve their connectivity. 
\item We optimize our proposed collaborative relaying approach to preserve the unbiasedness of the local updates and to minimize the expected convergence time to an optimal model.
\item Our approach can be applied to PSs that are blind to the identities of the transmitting clients. Therefore, it is suitable to be used in conjunction with  OAC.
\item  By conducting extensive simulations, we numerically show the superiority of the proposed framework compared to federated averaging (FedAvg), particularly  when the data heterogeneity is taken into account to mimic  realistic scenarios.
\end{itemize}

\subsection{Paper Organization}
The rest of the paper is organized as follows: Section \ref{sec:system_model} presents the FL system model and the proposed collaborative relaying scheme.
Section \ref{sec:unbiasedness_exp_dist_optimality}  derives conditions for the unbiasedness of  our collaborative relaying scheme and presents  an upper bound on its expected distance to optimality.
Using these analytical guarantees, Section \ref{sec:optimizing_alphas} optimizes our collaborative relaying scheme to reduce the upper bound on the expected suboptimality gap. Section \ref{sec:simulation_results}
presents numerical results that validate our theoretical analysis and highlight the performance improvement in terms of the training accuracy  collaborative relaying provides. 
Finally, Section \ref{sec:conclusion} concludes this paper.

\section{System Model for Collaborative Relaying} \label{sec:system_model}
Consider $n$ clients each with its own local dataset collaborating with the help of a central PS \my{to train} a common model with $d$ parameters represented by the $d$-dimensional vector $\xv\in\Real^d$. The clients communicate periodically with the PS over intermittently connected links to minimize an empirical loss function we define below.

Let $\Lc(\xv, \zetav)$ be the loss evaluated for a model $\xv$ at data point $\zetav$.
Denote the local dataset of client $i$ by $\Zc_i$, and its local loss function by $f_i : \Real^d \times \Zc_i \to \Real$, where $f_i(\xv; \Zc_i) = \frac{1}{\lvert \Zc_i \rvert}\sum_{\zetav \in\Zc_i}\Lc(\xv; \zetav)$.
The PS aims to solve the following empirical risk minimization (ERM) problem:
\begin{equation}
    \xv^* = \argminimize_{\xv \in \Real^d}f(\xv) \triangleq \argminimize_{\xv \in \Real^d} \frac{1}{n}\sum_{i=1}^{n}f_i(\xv;\Zc_i).
\end{equation}
We assume uniform sizes for local datasets, i.e., $|\Zc_i|=|\Zc_j|,\:\forall i,j\in[n]$.

\subsection{FL with Local SGD at Clients}
We denote by $\nabla f_i(\xv)$ the gradient of the loss function at client $i$, i.e., $\nabla f_i(\xv)\triangleq \nabla f_i(\xv;\Zc_i)$. Additionally, let $g_i(\xv)$ denote a stochastic gradient of $f_i(\xv;\Zc_i)$.

FL consists of two stages, a local optimization stage where clients update their local models iteratively, and an aggregation stage where the PS aggregates the local updates it receives from the clients.
Let  $\Tcal$ be the \textit{period} of local averaging, i.e., the number of local iterations \textit{after} which the PS receives clients' estimates for the model parameters.
At the beginning of the $r^{th}$ \textit{round} of FL, the PS broadcasts the global model $\xv^{(r)}$ to the clients. 
For local iteration $k \in [0:\Tc]$ of the $r^{th}$ round, client $i$ applies the following local update rule:
\begin{flalign}\label{eq:client_update_weighted_FL}
\xv_i^{(r,k+1)} &= \xv_i^{(r,k)}-\eta_rg_i\left(\xv_i^{(r,k)}\right),
\end{flalign}
where $\eta_r$ is the local learning rate for round $r$, and we set $\xv_i^{(r,0)}=\xv^{(r)}$.

\subsection{Communication Model}
\begin{figure}
\vspace{-0.5cm}
\centering
\includegraphics[width=.55\linewidth]{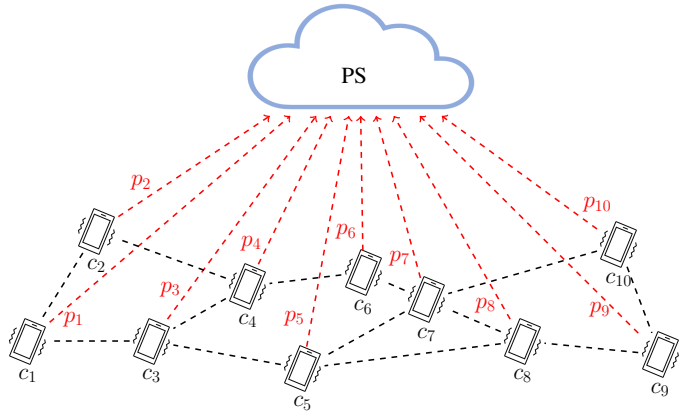}
\caption{Illustration of the system model. Red arrows denote intermittent uplink connections from the clients to the PS. 
Additionally, black dashed lines denote intermittent  connections between the clients.
Here, $c_i$ denotes client $i$, while $p_i$ is
the connectivity probability of the link from $c_i$ to the PS.
}
\label{fig:communication_model}
\end{figure}

For simplicity of the exposition, we consider communication links from the clients to the PS, where each link is either unavailable or perfect; that is, when it is available it does not suffer from any channel impairments. We depict the considered  communication model in Fig.~\ref{fig:communication_model}.

\subsubsection*{Communication between the clients and PS}
We model the intermittent connectivity of client $i$ to the PS at round $r$ by the Bernoulli random variable $\tau_i(r)\sim\text{Bernoulli}(p_i)$, where $\tau_i(r)\!=\!1$ denotes a communication opportunity between client $i$ and the PS at round $r$, whereas   $\tau_i(r)\!=\!0$ means the connection is blocked. 
We assume that the link connectivity to the PS is independent across rounds and across clients.

\subsubsection*{Communication between clients}
We consider intermittent connectivity for the links between the clients as well. We capture the intermittent connectivity of a transmission from client $i$ to client $j$ by the random variable $\tau_{ij}(r)\sim\text{Bernoulli}(p_{ij})$, where $p_{ii}\!=\!1$ for every $i\in[n]$.  For simplicity of exposition, we assume that the inter-client links and the links from the clients to the PS are statistically independent.

Additionally, we assume that  $\tau_{ij}(r)$, $\tau_{lm}(r')$, are statistically independent for every $i,j,l,m\in[n]$ and $r,r'\in\mathbb{N}$ such that $(i,j)\neq (l,m)$ and $(j,i)\neq (l,m)$, or $r\neq r'$.

We use the notation $E_{\{i,j\}} = \mathbb{E}[\tau_{ij}(r)\tau_{ji}(r)]$ for every $i,j\in[n]$ and $r\in\mathbb{N}$ to denote the correlation due to channel reciprocity. 
 Finally, we assume that $\tau_{i}(r+1), \tau_{i  j}(r+1)$ are statistically independent of $\Delta\xv_l^{r+1}$ for every $i,k,l\in[n]$

\textbf{Remark:}
 We assume that the connectivity probabilities $p_i,p_{ij},\:i,j\in[n]$, are known. In practice, they can be easily estimated. Moreover, clients can share their $p_i$ with each other using local links in a pre-training phase. On the other hand, we do not assume that the instantaneous connectivity information $\tau_i(r),\: r\in[n]$, is available to any of the clients.

We assume that the communicating clients can send their updates to  one another over the inter-client links. Therefore, each  client can send to the PS a weighted average of its own update and those of its neighbors.
In this way, the PS can receive  estimates from clients with  failing  uplink connections. Based on this observation, we next present our collaborative relaying procedure. 

\subsection{Collaborative Relaying of Local Updates}
Let $\Delta\xv_j^{r+1}$ denote the update of client $i$ at the end of the $\mathcal{T}$th local iteration in round $r$; i.e.,
\[\Delta\xv_j^{r+1}=\xv_j^{(r,\mathcal{T})}-\xv^{(r)}.\]
We assume that the model update of client $i$ at the end of the $\mathcal{T}$th iteration is readily available to every client $j$ such that $\tau_{i  j}(r+1)=1$.
Then, client $i$ forms a \my{linear} combination of its own update and those of its neighbors
 \begin{flalign}\label{eq:weighted_average_x_i}
     \Delta\widetilde{\xv}_i^{r+1}=   \sum_{j\in[n]} \tau_{j  i}(r+1)\alpha_{ij}\cdot\Delta\xv_j^{r+1} =   \sum_{j\in[n]} \alpha_{ij}\tau_{j  i}(r+1)\left(\xv_j^{(r,\mathcal{T})}-\xv^{(r)}\right),
 \end{flalign} 
 where $\alpha_{ij}$ is the weight used by client $i$ to relay the update from client $j$. The complexity of this averaging is low since it involves a weighted averaging of $O(\max_{i\in[n]}|\{j:p_{j  i}>0\}|+1)$.

\begin{algorithm}[t!]
\DontPrintSemicolon 
\KwIn{Round number $r$,  
step-size $\eta_r$,
round duration $\mathcal{T}$, neighbors of client $i$ $\mathcal{N}_i$, 
$\alpha_{ij}$ for every $j\in\mathcal{N}_i\cup\{i\}$.}
\KwOut{$\Delta\widetilde{\xv}_i^{r+1}$.}
Receive $\xv^{(r)}$ from  PS\;
Set $\xv_i^{(r,0)}=\xv^{(r)}$\;
\For{$k \gets 0$ \textbf{to} $\mathcal{T}-1$}{
  Generate a random gradient $g_i\left(\xv_i^{(r,k)}\right)$\; 
  $\xv_i^{(r,k+1)} = \xv_i^{(r,k)}-\eta_rg_i\left(\xv_i^{(r,k)}\right)$\;
  }
  Set  $\Delta\xv_i^{r+1}=\xv_i^{(r,\mathcal{T})}-\xv^{(r)}$\;
 Send $\Delta\xv_i$ to every $j\in\mathcal{N}_i$\;
 Receive $\Delta\xv_j$ from every $j\in[n]$ \my{such that $\tau_{j  i}(r+1)=1$}\;
 Compute  $\Delta\widetilde{\xv}_i^{r+1}=\sum_{j\in [n]} \tau_{j  i}(r+1)\alpha_{ij}\cdot\Delta\xv_j^{r+1}$\;
 Transmit $\Delta\widetilde{\xv}_i^{r+1}$ to the PS\;
 \caption{{\sc ColRel-Client} (Client Collaborative Relaying)} 
\label{algo:client_i_comp_forward}
\end{algorithm}

\subsection{PS Aggregation}
We assume that the PS does not explicitly select which subset of clients it will receive updates from, but rather receives information from all the \textit{communicating} clients in each round.

The PS can use the following rescaled sum of  the received updates:
\begin{flalign}\label{eq:averaging_update_weighted_FL}
\xv^{(r+1)} &= \xv^{(r)}+\frac{1}{n}\sum_{i\in[n]}\tau_i(r+1)
\Delta\widetilde{\xv}_i^{r+1}\nonumber\\
&= \xv^{(r)}+\frac{1}{n}\sum_{i\in[n]}\tau_i(r+1)\sum_{j\in[n]} \tau_{j  i}(r+1)\alpha_{ij}\left(\xv_j^{(r,\mathcal{T})}-\xv^{(r)}\right).
\end{flalign}
Note that this is a simple sum of the updates from all the connected devices, thus it  can be computed over-the-air and does not require the PS to know the identities of the communicating clients at each round.
We depict the overall collaborative relaying (ColRel)  FL procedure in Algorithm \ref{algo:client_i_comp_forward} and Algorithm \ref{algo:PS_comp_forward}.  Algorithm \ref{algo:client_i_comp_forward} describes the local updates at each client $i\in[n]$, and its collaborative relaying mechanism. Algorithm \ref{algo:PS_comp_forward} describes the PS aggregation.

\begin{algorithm}[t!]
\DontPrintSemicolon 
\KwIn{Number of rounds $R$,
a set of clients $[n]$, scalar $w$.}
\KwOut{An estimate $\xv^{(R)}$ of optimal value $\xv^{\star}$}
Set $\xv^{0}=\boldsymbol{0}$\;
\For{$r \gets 0$ \textbf{to} $R-1$}{
  Send $\xv^{(r)}$ to all clients\;
  Let $\tau_i(r+1)=1$ if  PS successfully receives $\Delta\widetilde{\xv}_i^{r+1}$ from client $i$ and set $\tau_i(r+1)=0$ otherwise.\;
  Compute $\xv^{(r+1)} = \xv^{(r)}+\frac{1}{n}\sum_{i\in[n]}\tau_i(r+1)
\Delta\widetilde{\xv}_i^{r+1}$
  }
\caption{{\sc ColRel-PS} (PS Aggregation)}
\label{algo:PS_comp_forward}
\end{algorithm}

\section{Unbiasedness and Expected Suboptimality Gap}\label{sec:unbiasedness_exp_dist_optimality} 
This section presents sufficient conditions for the unbiasedness of the proposed collaborative relaying method. Under the unbiasedness conditions, we derive an upper bound on the expected suboptimality gap of the proposed method.

\subsection{Sufficient Conditions for the Unbiasedness of Collaborative Relaying}

Recall that $\alpha_{ji}$ is the weight client $j$ assigns to the estimate it receives from client $i$.
 Client $i$ and each  client $j$ that receives the transmission from client $i$ successfully, i.e. $\{j\in[n]:\:\tau_{i  j}(r+1)=1\}$, try to send to the PS $\alpha_{ji}\Delta\xv_i^{r+1}$ on behalf of client $i$.
Then,  the  total received update  of client $i$ at the PS is given by 
\begin{flalign*}
\sum_{j\in [n]} \tau_j(r+1)\tau_{i  j}(r+1)\alpha_{ji}\Delta\xv_i^{r+1}.
\end{flalign*}
\my{We are interested in observing the objective function at the PS, thus, we aim at deriving a condition on the  weights $\alpha_{ji}$ such that the  function at the PS, which we want to minimize, remains unbiased.}
The following lemma presents a sufficient condition on $\alpha_{ji}$  for unbiasedness.

\begin{lemma}[Sufficient condition for unbiasedness]\label{lemma:unbiassed}
Let   $\alpha_{ij}$  be such that
\begin{align}\label{eq:alpha_sum_unbiassed_contraint}
\mathbb{E}\left[\sum_{j\in[n]} \tau_j(r+1)\tau_{i  j}(r+1)\alpha_{ji}\right] 
=p_i\alpha_{ii}+\sum_{j\neq i}p_jp_{i  j}\alpha_{ji}= 1.
\end{align}
Then, for every $i\in[n]$ and \my{$r\geq 0$}
\begin{flalign*}
\frac{1}{n} \cdot \mathbb{E}\left[\sum_{j\in [n]} \tau_j(r+1)\tau_{i  j}(r+1)\alpha_{ji}\Delta\xv_i^{r}\Big|\Delta\xv_i^{r+1}\right] =\frac{1}{n}\cdot
\Delta\xv_i^{r}.
\end{flalign*}
\end{lemma}
\begin{proof}
Since $\tau_j(r+1)$ and $\tau_{i  j}(r+1)$ are statistically independent of $\Delta\xv_i^{r}$ for all $i,j\in[n]$, we have that 
\begin{flalign*}
\mathbb{E}\left[\sum_{j\in[n]} \tau_j(r+1)\tau_{i  j}(r+1)\alpha_{ji}\Delta\xv_i^{r+1}\Big|\Delta\xv_i^{r+1}\right] =
\mathbb{E}\left[\sum_{j\in[n]} \tau_j(r+1)\tau_{i  j}(r+1)\alpha_{ji}\right] \Delta\xv_i^{r+1}.
\end{flalign*}
Substituting \eqref{eq:alpha_sum_unbiassed_contraint} concludes the proof.
\end{proof}

We note that if condition \eqref{eq:alpha_sum_unbiassed_contraint} is fulfilled, then we also have
\begin{flalign*}
\frac{1}{n}\cdot \mathbb{E}\left[\sum_{i\in[n]}\tau_i(r+1)\sum_{j\in [n]}\tau_{j  i}(r+1)\alpha_{ij}\right] &= \frac{1}{n}\sum_{i\in[n]}\mathbb{E}\left[\sum_{j\in [n]} \tau_j(r+1)\tau_{i  j}(r+1)\alpha_{ji}\right]=\frac{1}{n}\cdot n=1.
\end{flalign*}
However,  since $\tau_j(r+1)$ and $\tau_{j  i}(r+1)$ are   random variables  Lemma \ref{lemma:unbiassed} does \textit{not} imply that
\begin{flalign*}
\frac{1}{n}\cdot \sum_{j\in[n]} \tau_j(r+1)\tau_{i  j}(r+1)\alpha_{ji}\Delta\xv_i^{r+1} =\frac{1}{n}\cdot
\Delta\xv_i^{r+1}.
\end{flalign*}
Finally,  the standard model of FL with random client sampling and no connectivity among clients is captured by substituting   $p_{i  j}=0, p_i=p$, $\alpha_{ii}=1$, and $\alpha_{ij}=0$ for all $i,j\in[n]$ such that $j\neq i$.

\subsection{Expected Suboptimality Gap}\label{sec:exp_dist_optimality}
Next, we present an upper bound on the expected distance to optimality as a function of the weights $\alpha_{ij},\: i,j\in[n]$. 

\begin{assumption}\label{assumption:Lipschitz}
The loss functions $f_i$ are $L$-smooth
with respect to $\xv$. That is, for every $i\in[n]$ and $\xv,\yv\in\mathbb{R}^d$, we have 
$\|\nabla f_i(\xv)-\nabla f_i(\yv)\|\leq L\|\xv-\yv\|$.
\end{assumption}

\begin{assumption}\label{assumption:unbiased_bounded_stochastic_gradients}
The stochastic gradients $g_i(\xv)$ are unbiased and have bounded variance, i.e.:
\begin{enumerate}
\item $\mathbbm{E}[g_i(\xv)] = \my{\nabla f(\xv)}$ and
\item there exists $\sigma^2$ such that $\mathbbm{E}[\|g_i(\xv)-\nabla f_i(\xv)\|^2]\leq \sigma^2$ for every $i\in[n]$, $\xv\in\mathbb{R}^d$.
\end{enumerate}
\end{assumption}
\my{The main focus of this paper is mitigating the harmful effect of intermittent connectivity for the convergence of the global model at the PS. Thus, for simplicity of exposition, we present analytical results for homogeneous data distribution setting, where $E[g_i(\xv)] = \my{\nabla f(\xv)}$. Our numerical results show that the collaborative relaying approach and the proposed weight optimization algorithms we present in this paper lead to a significant reduction in the training loss in heterogeneous data settings as long as $\mathbbm{E}[g_i(\xv)] = \my{\nabla f_i(\xv)}$.}

\begin{assumption}\label{assumption:Mu_stongly_convex}
The loss functions $f_i$ are $\mu$ strongly convex. That is, for every $i\in[n]$ and $\xv,\yv\in\mathbb{R}^d$ we have
$
\langle\nabla f_i(\xv)-\nabla f_i(\yv),\xv-\yv\rangle\geq \mu\|\xv-\yv\|^2$.

\end{assumption}

\begin{assumption}\label{assumption:positive_alpha}
The weights $\alpha_{ij}$ are nonnegative, i.e.,  $\alpha_{ij}\geq0,\: \forall \:i,j\in[n]$.
\end{assumption}

Before stating the main result, we introduce some definitions. We let $\boldsymbol\Alpha\triangleq(\alpha_{ij})_{i,j\in [n]}$ and $\boldsymbol{P}\triangleq(p_{i  j})_{i,j\in [n]}$.  Additionally, we define 
\begin{flalign*}
S(\boldsymbol{p},\boldsymbol{P}, \boldsymbol\Alpha) \triangleq & \sum_{i,j,l\in[n]}
 p_j(1-p_j)p_{i  j}p_{l  j}\alpha_{ji}\alpha_{jl}+\sum_{i,j\in[n]}p_{i  j}p_j(1-p_{i  j})\alpha^2_{ji}+\sum_{i,l\in[n]}p_ip_l(E_{\{i,l\}}-p_{i  l}p_{l  i})\alpha_{il}\alpha_{li}.
\end{flalign*}

\begin{theorem}[Expected Distance to Optimality]\label{theorem:expected_distance_to_opt}
Let
\begin{flalign*}
B(\boldsymbol{p},\boldsymbol{P}, \boldsymbol\Alpha) \triangleq\frac{2L^2}{n^2}S(\boldsymbol{p},\boldsymbol{P}, \boldsymbol\Alpha) 
\quad\text{ and }\quad
r_0(\boldsymbol{p},\boldsymbol{P}, \boldsymbol\Alpha)\triangleq \max\left\{\frac{L}{\mu},4\left(\frac{B(\boldsymbol{p},\boldsymbol{P}, \boldsymbol\Alpha) }{\mu^2}+1\right),\frac{1}{\mathcal{T}},\frac{4n}{\mu^2\mathcal{T}}\right\}.
\end{flalign*}
Additionally, let $\eta_r\triangleq\frac{4\mu^{-1}}{r\mathcal{T}+1}$, $ C_1(\boldsymbol{p},\boldsymbol{P}, \boldsymbol\Alpha) \triangleq \frac{4^2}{\mu^2}\cdot \frac{2\sigma^2}{n^2}S(\boldsymbol{p},\boldsymbol{P}, \boldsymbol\Alpha)$, $C_2 \triangleq \frac{4^2}{\mu^2}\cdot L^2\frac{\sigma^2}{n}e$, and $C_3(\boldsymbol{p},\boldsymbol{P}, \boldsymbol\Alpha) \triangleq \frac{4^4}{\mu^4}\cdot\left( L^2\sigma^2e+\frac{2L^2\sigma^2e}{n^2}S(\boldsymbol{p},\boldsymbol{P}, \boldsymbol\Alpha)\right)$.
Then, under Assumptions \ref{assumption:Lipschitz}-\ref{assumption:positive_alpha} and condition \eqref{eq:alpha_sum_unbiassed_contraint}, 
for every $r\geq r_0(\boldsymbol{p},\boldsymbol{P}, \boldsymbol\Alpha)$ we have 
\begin{flalign}\label{eq:expected_distance_to_opt}
&\mathbbm{E}\left\|\xv^{(r+1)}-x^{\star}\right\|^2\leq\nonumber\\
&\frac{(r_0\mathcal{T}+1)}{(r\mathcal{T}+1)^2}\left\|\xv^{(0)}-x^{\star}\right\|^2+ C_1(\boldsymbol{p},\boldsymbol{P}, \boldsymbol\Alpha)\frac{\mathcal{T}}{r\mathcal{T}+1}+C_2\frac{(\mathcal{T}-1)^2}{r\mathcal{T}+1}+C_3(\boldsymbol{p},\boldsymbol{P}, \boldsymbol\Alpha)\frac{\mathcal{T}\my{-1}}{(r\mathcal{T}+1)^2}, 
\end{flalign}
given the update dynamic captured by \eqref{eq:client_update_weighted_FL}-\eqref{eq:averaging_update_weighted_FL}.
\end{theorem} 
\my{We present the proof of Theorem \ref{theorem:expected_distance_to_opt} in Appendix \ref{appnd:proof:theorem_convergence_rate}.} 

Theorem \ref{theorem:expected_distance_to_opt} implies that  
\begin{flalign*}
E\left\|\xv^{(r+1)}-x^{\star}\right\|^2=O\left(\frac{\left\|\xv^{(0)}-x^{\star}\right\|^2}{r^2}+\frac{S(\boldsymbol{p},\boldsymbol{P}, \boldsymbol\Alpha)}{r}\right).
\end{flalign*}
Consequently, we can reduce the upper bound on the distance to optimality \eqref{eq:expected_distance_to_opt} by minimizing the term $S(\boldsymbol{p},\boldsymbol{P},\boldsymbol\Alpha)$ under the unbiasedness condition  \eqref{eq:alpha_sum_unbiassed_contraint}. 

\section{Optimizing the Weights $\alpha_{ij}$}\label{sec:optimizing_alphas}
Based on Theorem \ref{theorem:expected_distance_to_opt} 
we can minimize the upper bound on the expected distance to optimality by solving the following optimization problem
\begin{flalign}\label{eq:minimize_distance}
&\min_{\boldsymbol\Alpha} S(\boldsymbol{p},\boldsymbol{P}, \boldsymbol\Alpha)\nonumber\\
&\text{s.t.: } \sum_{j\in [n]}p_jp_{i  j}\alpha_{ji}= 1,\qquad  \alpha_{ji}\geq 0\quad \forall i,j\in[n].
\end{flalign}
Due to the term $\sum_{i,l\in[n]}p_ip_l(E_{\{i,l\}}\hspace{-0.04cm}-\hspace{-0.04cm}p_{i  l}p_{l  i})\alpha_{il}\alpha_{li}$, the function $S(\boldsymbol{p},\boldsymbol{P}, \boldsymbol\Alpha)$ is not necessarily convex with respect to $\boldsymbol\Alpha$. Therefore, we will instead consider the convex upper bound $\overline{S}(\boldsymbol{p},\boldsymbol{P}, \boldsymbol\Alpha)$ where,
\begin{flalign*}
\overline{S}(\boldsymbol{p},\boldsymbol{P}, \boldsymbol\Alpha)\triangleq  \sum_{i,j,l\in[n]}
 p_j(1-p_j)p_{i  j}p_{l  j}\alpha_{ji}\alpha_{jl}+\sum_{i,j\in[n]}p_{i  j}p_j(1-p_{i  j})\alpha^2_{ji}+\sum_{i,l\in[n]}p_ip_l(E_{\{i,l\}}-p_{i  l}p_{l  i})\alpha^2_{li}.
\end{flalign*}

Recalling that $E_{\{i,j\}}\geq p_{ij}p_{ji}$, $\forall\: i,j\in[n],$ we show in Lemma \ref{lemma:conex_S_func_A} that minimizing $\overline{S}(\boldsymbol{p}, \boldsymbol{P}, \boldsymbol\Alpha)$ is a convex relaxation of \eqref{eq:minimize_distance}. 
\my{Thus}, we initially use the Gauss-Seidel method to iteratively minimize $\overline{S}(\boldsymbol{p},\boldsymbol{P}, \boldsymbol\Alpha)$ with respect to $\boldsymbol\Alpha$, we denote this solution by $\boldsymbol\Alpha^*$. 
Then, we use the final iterate as a warm-start initialization for Gauss-Seidel method to iteratively converge to a stationary point of $S(\boldsymbol{p},\boldsymbol{P}, \boldsymbol\Alpha)$ in the vicinity of $\boldsymbol\Alpha^*$.

\begin{lemma}\label{lemma:conex_S_func_A}
For every $\boldsymbol\Alpha$ such that $\alpha_{ij}\geq0,\:\forall\: i,j\in[n]$,  $S(\boldsymbol{p},\boldsymbol{P}, \boldsymbol\Alpha)\leq\overline{S}(\boldsymbol{p},\boldsymbol{P}, \boldsymbol\Alpha)$.
Additionally, 
$\overline{S}(\boldsymbol{p}, \boldsymbol{P}, \boldsymbol\Alpha)$ is convex with respect $\boldsymbol\Alpha$ for $\boldsymbol{p}\in[0,1]^{n}$.
\end{lemma}
We present the proof of this lemma in Appendix \ref{append:proof_lemma_conex_S_func_A}.
By Lemma \ref{lemma:conex_S_func_A},  the following optimization problem is convex:
\vspace{-0.3cm}
\begin{flalign}\label{eq:minimize_distance_upper}
&\min_{\boldsymbol\Alpha} \overline{S}(\boldsymbol{p},\boldsymbol{P}, \boldsymbol\Alpha)\nonumber\\
\text{s.t.:}&\sum_{j\in[n]} p_jp_{i  j}\alpha_{ji}=1,\quad \;\; \alpha_{ji}\geq 0\quad \forall i,j\in[n].
\end{flalign}

Let $\boldsymbol\Alpha_{i}$ denote the $i$th column of $\boldsymbol\Alpha$, that is, $\boldsymbol\Alpha_{i}=\left(\boldsymbol\Alpha_{1i},\ldots,\boldsymbol\Alpha_{ni}\right)^T$. 
Since the domain of the problems \eqref{eq:minimize_distance} and \eqref{eq:minimize_distance_upper} is separable with respect to  $\boldsymbol\Alpha_{i}$, we can use the Gauss-Seidel method to  iteratively solve \eqref{eq:minimize_distance_upper} and converge to an optimal solution \cite[Proposition 2.7.1]{Bertsekas_nonlinear_programming} for the upper bound \eqref{eq:minimize_distance_upper}. We can then converge to a stationary point of \eqref{eq:minimize_distance} in the vicinity of this solution by utilizing again the  Gauss-Seidel method.

\textbf{Remark:}
We note that when $p_{i  j}\in\{0,1\},\:\forall i,j\in [n]$ the problem \eqref{eq:minimize_distance} is convex. Additionally, in this case the problems \eqref{eq:minimize_distance} and \eqref{eq:minimize_distance_upper} coincide since $E_{\{i,l\}}-p_{i  l}p_{l  i}=0$ for every $i,j\in [n]$.

\paragraph{Optimizing the convex relaxation \eqref{eq:minimize_distance_upper}}
Let  $\boldsymbol{\Alpha}_i^{(\ell)}$ denote the approximated value for $\boldsymbol{\Alpha}_{i}$ in the $\ell$th iteration of the Gauss-Seidel method of \eqref{eq:minimize_distance_upper}.
We choose the initial solution $\boldsymbol{\Alpha}_{ji}^{(0)}=\frac{1}{\left(\sum_{k\in[n]}\mathbbm{1}_{\{p_k>0,p_{i  k}>0\}}\right)} p_{j}p_{i  j}\cdot\mathbbm{1}_{\{p_j>0,p_{i  j}>0\}}$. 
We can improve our solution iteratively by the Gauss-Seidel method until convergence to an optimal point of \eqref{eq:minimize_distance_upper}. 
That is, at every  iteration $\ell$ we compute $\boldsymbol{\Alpha}^{(\ell)}$ as follows
\begin{flalign}\label{eq:A_hat_iteration_ell_upper}
\boldsymbol{\Alpha}_i^{(\ell)}=\begin{cases}
\widehat{\boldsymbol{\Alpha}}_i^{(\ell)} & \text{\: if \:} \ell\mod{n}+n\cdot\mathbbm{1}_{\{\ell\mod{n}=0\}}=i\\
\boldsymbol{\Alpha}_i^{(\ell-1)}& \text{\: otherwise},
\end{cases}
\end{flalign}
where 
\begin{flalign}\label{eq:opt_alpha_allocation_iterative_upper}
\widehat{\boldsymbol\Alpha}_{i}^{(\ell)}&=\arg\min\left[\sum_{j\in[n]}p_jp_{i  j}\left(1-p_jp_{i  j}\right)\alpha_{ji}^2+2\sum_{l\in[n],l\neq i}\sum_{j\in[n]} p_j(1-p_j)p_{i  j}p_{l  j}\alpha_{ji}\alpha_{jl}^{(\ell-1)}\right.\nonumber\\
&\hspace{7cm}\left.+\sum_{j\in[n]}p_ip_j(E_{\{i,j\}}-p_{i  j}p_{j  i})\alpha^2_{ji}\right],\nonumber\\
&\quad\text{s.t.:}\sum_{j\in[n]} p_jp_{i  j}\alpha_{ji}=1, \;\;  \alpha_{ji}\geq 0,\quad \forall j\in[n].
\end{flalign}

Using Lagrange multipliers, we show in Appendix \ref{append:Solve_alpha_allocation_lagrange} that the optimal value for $\widehat{\boldsymbol\Alpha}_i^{(\ell)}$ is:
\begin{flalign}\label{eq:alpha_i_clac_cur_upper}
\widehat{\alpha}_{ji}^{(\ell)}(\lambda_i) = 
\begin{cases}
\left(\frac{-2(1-p_j)\sum_{l\in [n]:l\neq i} p_{l  j}\alpha_{jl}^{(\ell-1)}+\lambda_i }{2[\left(1-p_jp_{i  j}\right)+p_i(E_{\{i,j\}}/p_{i  j}-p_{j  i})]}\right)^+&
\text{ if } p_jp_{i  j}\cdot\max_{k\in[n]}p_kp_{i  k}\in(0,1),\\
\frac{1}{\sum_{k\in[n]}\mathbbm{1}_{\{p_kp_{i  k}=1\}}} & \text{ if } p_jp_{i  j}=1,\\
0 & \text{ otherwise},
\end{cases}
\end{flalign}
where $(a)^+\triangleq\max\{a,0\}$ and
$\lambda_i$ is set such that $\sum_{j\in[n]}p_jp_{i  j}\widehat{\alpha}_{ji}^{(\ell)}(\lambda_i)=1$. We can find  $\lambda_i$ using the bisection method over the interval 
$\big[0,\max_{j:p_jp_{i  j\in(0,1)}}\big\{\frac{2[\left(1-p_jp_{i  j}\right)+p_i(E_{\{i,j\}}/p_{i  j}-p_{j  i})]}{p_jp_{i  j}}\allowbreak +2(1-p_j)\sum_{l\in [n]:l\neq i} p_{l  j}\alpha_{jl}^{(\ell-1)}\big\}\big]$

\paragraph{Fine tuning \eqref{eq:minimize_distance}} 
Let  $\boldsymbol{\Alpha}_i^{(\ell)}$ denote the approximated value for $\boldsymbol{\Alpha}_{i}$ in the $\ell$th iteration of the Gauss-Seidel method of \eqref{eq:minimize_distance}, and assume a given initialization $\boldsymbol{\Alpha}_i^{(0)}$. We can improve our solution iteratively by the Gauss-Seidel method until convergence to a stationary point of \eqref{eq:minimize_distance}. 

At every  iteration $\ell$ we compute $\boldsymbol{\Alpha}^{(\ell)}$ as follows
\begin{flalign}\label{eq:A_hat_iteration_ell}
\boldsymbol{\Alpha}_i^{(\ell)}=\begin{cases}
\widehat{\boldsymbol{\Alpha}}_i^{(\ell)} & \text{\: if \:} \ell\mod{n}+n\cdot\mathbbm{1}_{\{\ell\mod{n}=0\}}=i\\
\boldsymbol{\Alpha}_i^{(\ell-1)}& \text{\: otherwise},
\end{cases}
\end{flalign}
where $\mathbbm{1}_{\{ \cdot \}}$ denotes the indicator function, and
\begin{flalign}\label{eq:opt_alpha_allocation_iterative}
\widehat{\boldsymbol\Alpha}_{i}^{(\ell)}&=\arg\min\left[\sum_{j\in[n]}p_jp_{i  j}\left(1-p_jp_{i  j}\right)\alpha_{ji}^2+2\sum_{l\in[n],l\neq i}\sum_{j\in[n]} p_j(1-p_j)p_{i  j}p_{l  j}\alpha_{ji}\alpha_{jl}^{(\ell-1)}\right.\nonumber\\
&\hspace{7cm}\left.+2\sum_{j\in[n]}p_ip_j(E_{\{i,j\}}-p_{i  j}p_{j  i})\alpha_{ij}\alpha_{ji}\right],\nonumber\\
\text{s.t.:}&\sum_{j\in[n]} p_jp_{i  j}\alpha_{ji}=1, \;\; \alpha_{ji}\geq 0,\quad \forall j\in[n].
\end{flalign}
Using Lagrange multipliers, we show in Appendix \ref{append:Solve_alpha_allocation_lagrange} that the optimal value for $\widehat{\boldsymbol\Alpha}_i^{(\ell)}$ is as follows:
\begin{flalign}\label{eq:alpha_i_clac_cur}
\widehat{\alpha}_{ji}^{(\ell)}(\lambda_i) = 
\begin{cases}
\left(\frac{-2(1-p_j)\hspace{-0.2cm}\sum\limits_{l\in [n]:l\neq i} p_{l  j}\alpha_{jl}^{(\ell-1)}-2p_i\big(\frac{E_{\{i,j\}}}{p_{i  j}}-p_{j  i}\big)\alpha_{ij}^{(\ell-1)}+\lambda_i }{2\left(1-p_jp_{i  j}\right)}\right)^+&
\text{ if } p_jp_{i  j}\max\limits_{k\in[n]}p_kp_{i  k}\in(0,1),\\
\frac{1}{\sum_{k\in[n]}\mathbbm{1}_{\{p_kp_{i  k}=1\}}} & \text{ if } p_jp_{i  j}=1,\\
0 & \text{ otherwise},
\end{cases}
\end{flalign}
where 
$\lambda_i$ is set such that $\sum_{j\in[n]}p_jp_{i  j}\widehat{\alpha}_{ji}^{(\ell)}(\lambda_i)=1$. We can find $\lambda_i$ using the bisection method over the interval 
$\big[0,\max_{j:p_jp_{i  j\in(0,1)}}\big\{\frac{2\left(1-p_jp_{i  j}\right)}{p_jp_{i  j}}+2(1-p_j)\sum_{l\in [n]:l\neq i} p_{l  j}\alpha_{jl}^{(\ell-1)}\allowbreak+2p_i\big(\frac{E_{\{i,j\}}}{p_{i  j}}-p_{j  i}\big)\alpha_{ij}^{(\ell-1)}\big\}\big]$.

We summarize the centralized optimization procedure for  $\boldsymbol{\Alpha}$ in Algorithm \ref{algo:opt_alpha_centralized}. 
 Additionally,  when the connectivity between two communicating clients is reliable, i.e., two clients are either connected with high probability or disconnected,  Algorithm \ref{algo:opt_alpha_centralized} can be implemented in a distributed fashion by the clients. In this case each client does not need to fully know  $\boldsymbol{p},\boldsymbol{P}$ and $\boldsymbol{\Alpha}$, but only the weights and transmission probabilities of all its direct neighbors and its second degree neighbors (i.e., neighbors of its neighbors). Such a distributed algorithm can be used to optimize the weights when the PS is blind to the identities of the transmitting clients at all times and cannot use a training period for learning the transmission probabilities. Finally, we note the weights resulting from Algorithm  \ref{algo:opt_alpha_centralized} can be used  as long as the probabilities $\boldsymbol{p}$ and $\boldsymbol{P}$ are fixed, and are not needed to be calculated in every communication round.

\paragraph*{Computation complexity} The overall computational complexity of Algorithm \ref{algo:opt_alpha_centralized}  is $O(I\cdot( n^2+K))$, where $K$ is the number of iterations used in the bisection method for optimizing $\lambda_i$. 

\begin{algorithm}[t!]
\DontPrintSemicolon 
\KwIn{A set of clients $[n]$, a connectivity graph $G$, the functions $\overline{S}(\boldsymbol{p},\boldsymbol{P},\boldsymbol\Alpha)$ and $S(\boldsymbol{p},\boldsymbol{P},\boldsymbol\Alpha)$, vector of connectivity probabilities $\boldsymbol{p}$, matrix of inter-client connectivity probabilities $\boldsymbol{P}$, maximal number of iteration $I$.} 
\KwOut{A matrix $\boldsymbol{\Alpha}^{(L)}$ that approximately minimizes \eqref{eq:minimize_distance}}
 \textbf{Initialize (convex relaxation)}:  $\boldsymbol\Alpha_{ji}^{(0)}=\frac{1}{\left(\sum_{j\in[n]}\mathbbm{1}_{\{p_j>0,p_{i  j}>0\}}\right) p_{j}p_{i  j}}\cdot\mathbbm{1}_{\{p_j>0,p_{i  j}>0\}}$, and $\ell=0$. \;
\While{$\ell\leq I-1$}{
 $\ell \leftarrow \ell+1$\;
$i \leftarrow \ell\mod{n}+n\cdot\mathbbm{1}_{\{\ell\mod{n}=0\}}$\;
Compute $\widehat{\boldsymbol\Alpha}_i^{(\ell)}$ according to \eqref{eq:alpha_i_clac_cur_upper}\;
 Set $\boldsymbol\Alpha_k^{(\ell)}$ according to \eqref{eq:A_hat_iteration_ell_upper}  for every $k\in[n]$\;
}
 \textbf{Warm-initialize (fine tuning)}: $\boldsymbol\Alpha^{(0)}=\boldsymbol\Alpha^{(L)}$, and $\ell=0$\;
\While{$\ell\leq I-1$}{
 $\ell \leftarrow \ell+1$\;
$i \leftarrow \ell\mod{n}+n\cdot\mathbbm{1}_{\{\ell\mod{n}=0\}}$\;
Compute $\widehat{\boldsymbol\Alpha}_i^{(\ell)}$ according to \eqref{eq:alpha_i_clac_cur}\;
 Set $\boldsymbol\Alpha_k^{(\ell)}$ according to \eqref{eq:A_hat_iteration_ell}  for every $k\in[n]$\;
}
\caption{{\sc COPT-$\alpha$} Centralized optimization of the weight matrix $\boldsymbol{\Alpha}$}  
\label{algo:opt_alpha_centralized}
\end{algorithm}

\section{Numerical Simulations}\label{sec:simulation_results}
\paragraph*{\textbf{Simulation setup}}
We train a ResNet-$20$ model for image classification on the CIFAR-10 \cite{cifar10} dataset.
We distributed the training set of $50,000$ images across $n\! =\!10$ clients in both independent and identically distributed (IID) and non-IID fashions.
Non-IID data distribution amongst clients is prevalent in FL applications,  to emulate it, we consider the \textit{sort-and-partition} approach wherein the training data is initially sorted based on the labels, and then they are divided into blocks and distributed among the clients randomly based on a parameter $s$, which measures the skewness of the data distribution.
More precisely, $s$ defines the maximum number of different
labels present in the local dataset of each user, and therefore, smaller $s$ implies a more skewed data distribution. 
We use $s = 3$, i.e. each client has images from at most 3 classes.

The plotted results for all the simulations are obtained by averaging over $5$ independent realizations.
In between every communication round to the PS, the clients execute $8$ local training steps of local-SGD.
We utilize the SGD optimizer at the clients with a global momentum $(\beta = 0.9)$ at the PS.
We used a learning rate of $0.05$ for SGD, a coefficient of $10^{-4}$ for $\ell_2$-regularization to prevent overfitting, and a batch-size of $64$.
All simulations were carried out on NVIDIA GeForce GTX $1080$ Ti with a CUDA Version $11.4$.
We also note that we have ensured all the simulations to have the same step-size for a fair comparison.
In other words, we have \textit{not} optimized the learning-rate for individual simulations.

To illustrate the advantages of \textsc{ColRel}
we compare it with three benchmark strategies:\\
 \textbf{FedAvg -- perfect connectivity}. 
    We consider FedAvg when all the clients are able to successfully transmit their local updates to the PS at every communication round. This serves as a natural upper bound to the performance of any algorithm proposed in the presence of intermittent connectivity.\\
\textbf{FedAvg -- Blind}.
     As a natural performance lower bound in the presence of intermittent client connectivity, we consider a na\"ive FedAvg strategy wherein the PS is unaware of the identity of clients.
     In this strategy, for the clients that are unable to send their updates to the PS due to a communication failure, the PS simply assumes that their update is zero.
     Essentially, the PS adds all the local updates it receives at any communication round, and divides it by the total number of clients irrespective of the knowledge of the number of actual successful transmissions. 
     Such blind averaging strategies are often the norm for FEEL employing OAC.\\
  \textbf{FedAvg -- Non-Blind}.
     As another benchmark, we  consider a non-blind strategy, where the PS is aware of the identity of the clients, and knows exactly, how many and which clients have successfully been able to send their local update to the PS. 
     This is common in point-to-point learning settings. 
     In this case, the PS simply ignores the clients that have been unable to send their updates, and averages the successful updates by dividing the global aggregate at the PS by the number of successful transmissions.

We compare the above-mentioned strategies with our proposed collaborative relaying strategy in the presence of intermittent connectivity of clients to the PS, as well as amongst themselves.
In order to demonstrate the improved performance of our proposed strategy with respect to the above-mentioned benchmarks, we consider the following setups:\\
 \textbf{Effect of topology}.
    The decentralized topology according to which the clients collaborate in sharing their updates plays an important role in the performance of \textsc{ColRel}.
    In particular, we investigate a setting in which intermittent connectivity between clients might be preferred over perfect connectivity because it leads to  increased collaboration on average.\\
 \textbf{Non-IID data distribution}.
    We evaluate the performance of \textsc{ColRel} in the presence of non-IID local data distribution.
    For instance, \textsc{ColRel} outperforms other strategies when some clients with important local data have persistently poor connectivity.\\
 \textbf{Heterogeneity in client connectivity}.
    Different clients may have different connectivity to the PS, and we show that as long as there is one client with reasonably good connection to the PS, its neighbors can relay their updates to it -- yielding improved performance for \textsc{ColRel}.

\subsubsection{IID local data distribution across clients}

\begin{figure}[t!]
\hspace{-0.3cm}
    \begin{subfigure}[t]{.48\textwidth}
    \centering
    \includegraphics[width=1.1\linewidth,trim={2cm 0.8cm 2.5cm 1.8cm},clip]{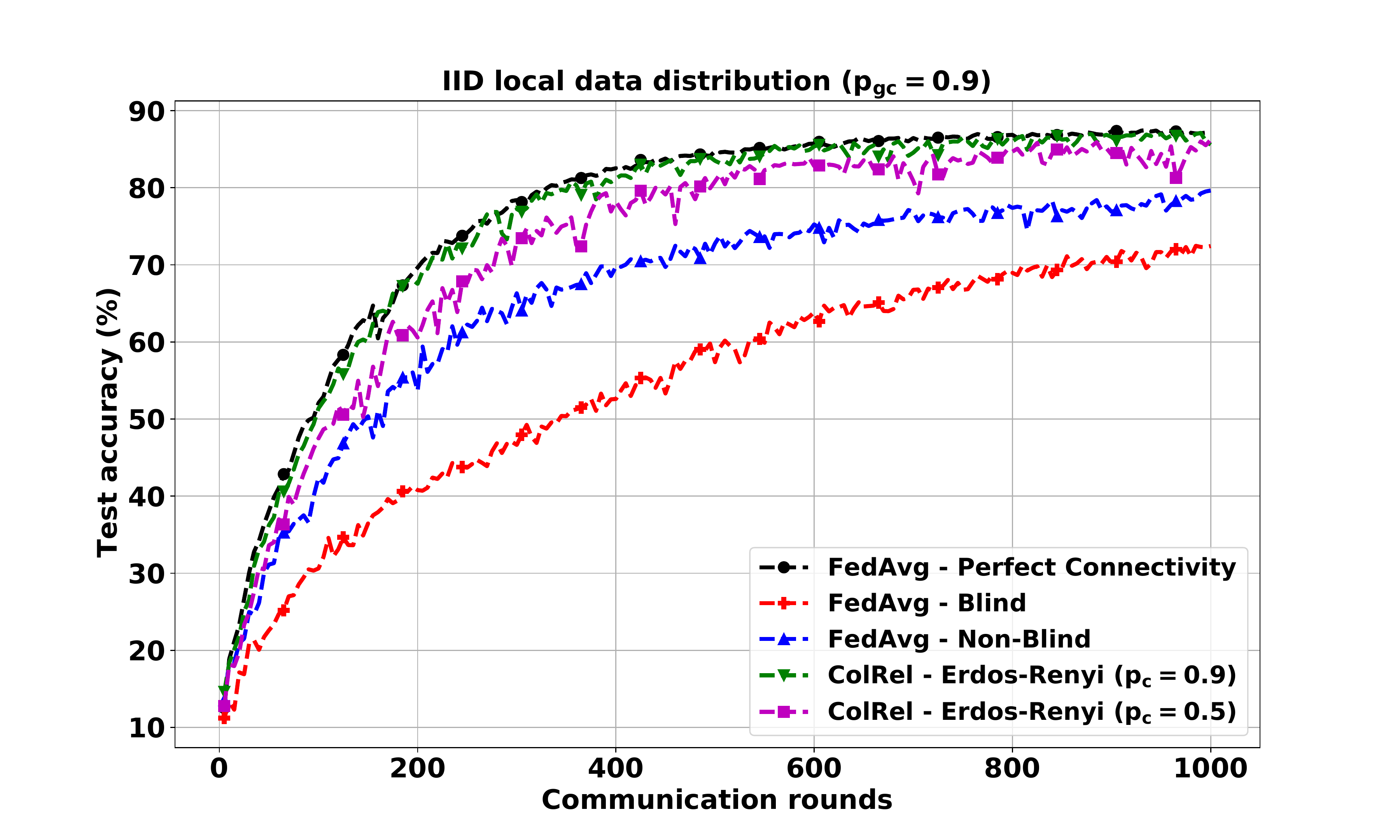}
    \caption{Comparison of federated learning strategies with only one client with good connectivity to the PS}
    \label{fig:int_iid_fully_connected}
  \end{subfigure}
  \hspace{1cm}
  \begin{subfigure}[t]{.48\textwidth}
    \centering
    \includegraphics[width=1.1\linewidth,trim={2cm 0.8cm 2.5cm 1.8cm},clip]{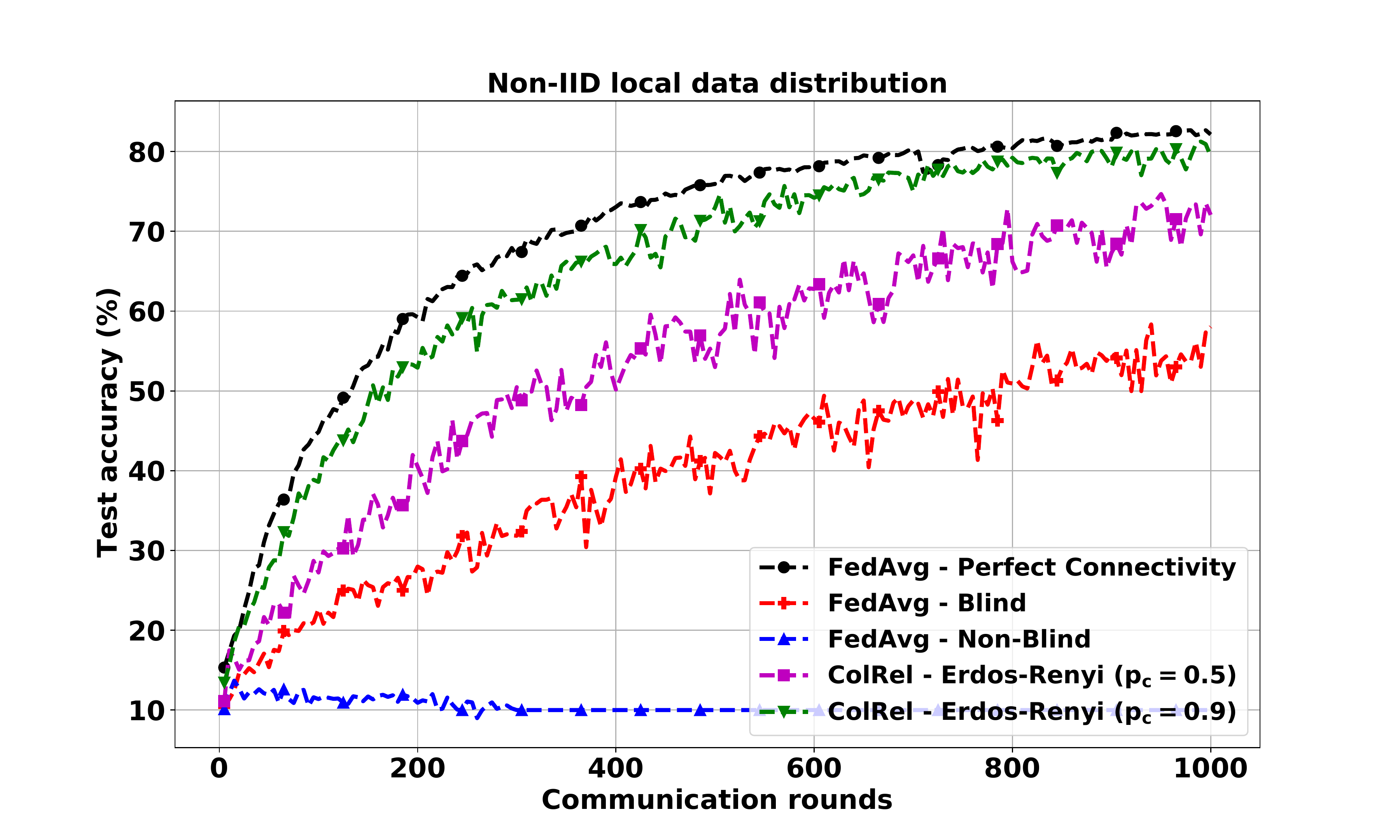}
    \caption{Comparison of federated learning strategies with heterogenous connectivity of clients to the PS}
  \label{fig:int_noniid_fully_connected}
  \end{subfigure}
  \vspace{4mm}
  \caption{Network topologies for perfect vs. intermittent client-client connectivities}
\end{figure}

In Fig. \ref{fig:int_iid_fully_connected}, we consider the setup wherein only one client has good connectivity to the PS with probability $p_{gc} = 0.9$, while all the remaining clients have a probability of only $0.1$ for successfully transmitting their updates to the PS.
For collaboration, we consider an Erd\H{o}s-R\'{e}nyi topology wherein a connection between any two clients is present with a certain probability $p_{c}$, with an additional constraint that $\tau_{ij} = 0 \iff \tau_{ji} = 0$.
The cases $p_{c} = 0.9$ and $p_{c} = 0.5$ correspond to different degrees of intermittent decentralized collaboration.
The performance of our proposed collaborative strategy, \textsc{ColRel} is comparable to FedAvg with perfect connectivity.
We also note that we have ensured all the simulations to have the same step-size, i.e. the learning rate for different simulations has not been tuned individually.

\subsubsection{Non-IID local data distribution across clients}
In Fig. \ref{fig:int_noniid_fully_connected}, we consider different clients have different connection probabilities to the PS.
In particular, some clients have a low probability of transmission, namely $p_1 = p_4 = p_5 = p_8 = 0.1$, some others moderate, and a couple of them high, i.e., $p_7 = 0.8$ and $p_{10} = 0.9$.
Once again, \textsc{ColRel} for both cases when $p_c = 0.9$ and $p_c = 0.5$ outperform both blind and non-blind FedAvg algorithms, even with the relatively low local connectivity with $p_c = 0.5$. we also observe that the convergence of \textsc{ColRel} is faster and more stable when the  inter-client connectivity probability $p_c$ increases.

\subsubsection{Network topology with mmWave links}
In our preliminary work \cite{our_ISIT_version}, we allow cooperation between clients only if they are connected via permanent communication links,  that is, two clients $i$ and $j$ are connected so as to allow collaboration if they are within each other's coverage area. Thus, \cite{our_ISIT_version}  assumes that $p_{ij}$ and $p_{ji}$ are either $0$ or $1$.
In this work, we allow for client-client cooperation even when the communication links between them are intermittent. 

Let $d_i$ be the distance between client $i$ an the PS and let $d_{ij}$ be the distance between client $i$ and client $j$, in meters.
We  consider a network topology with mmWave links, as in \cite{6834753}, in which 
$p_{i} = \min\left(1, \mathrm{e}^{-d_{i}/30 + 5.2}\right)$ and 
$p_{ij} = \min\left(1, \mathrm{e}^{-d_{ij}/30 + 5.2}\right)$.

\begin{figure}[t!]
    \begin{subfigure}[t]{.48\textwidth}
    \centering
    \includegraphics[width=0.7\linewidth, trim={0.5cm 0.5cm 0.2cm 0.15cm},clip]{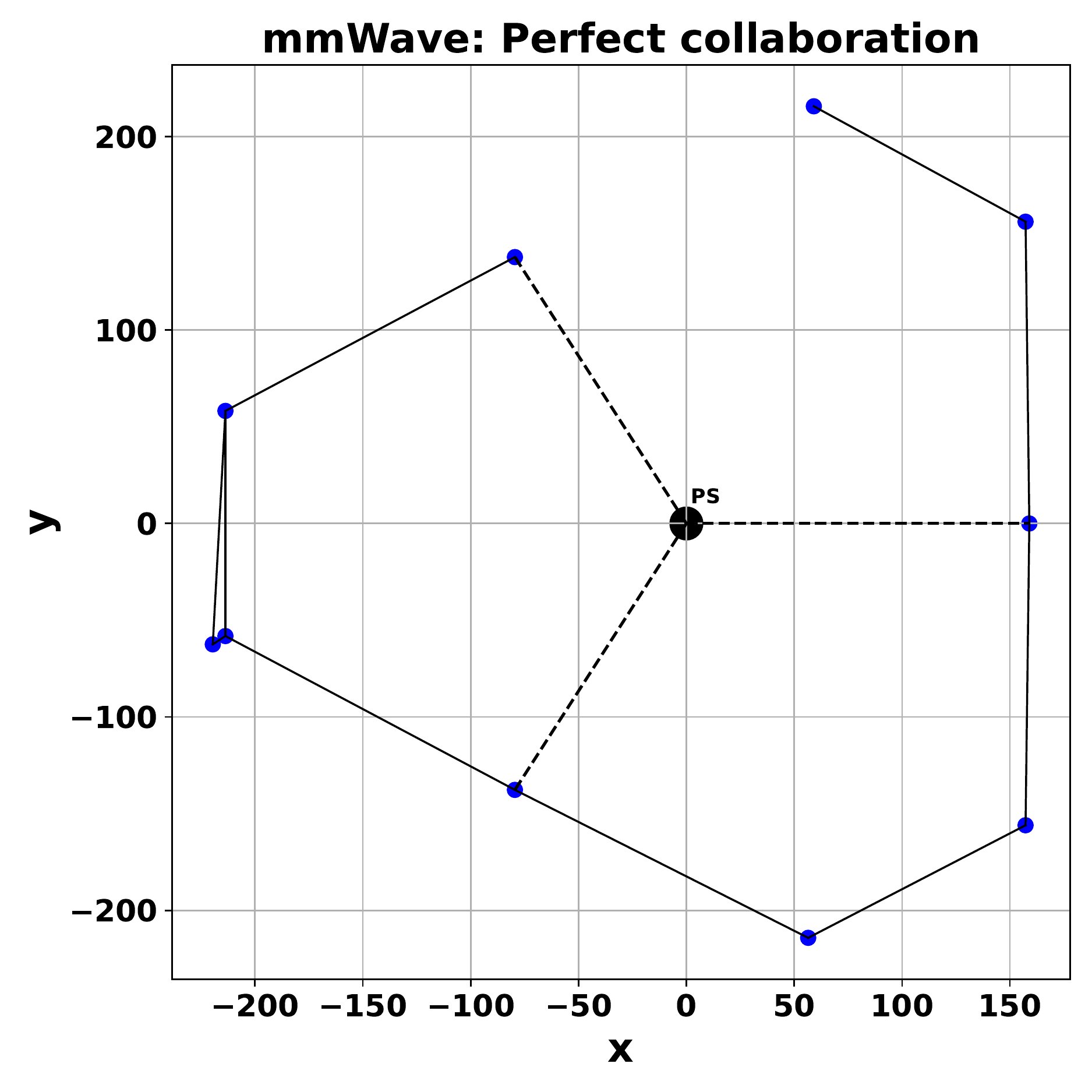}
    \caption{Perfect client-client collaboration}
    \label{fig:mmWave_perfect_topology}
  \end{subfigure}
  \hfill
  \begin{subfigure}[t]{.48\textwidth}
    \centering
    \includegraphics[width=0.7\linewidth, trim={0.5cm 0.5cm 0.2cm 0.15cm},clip]{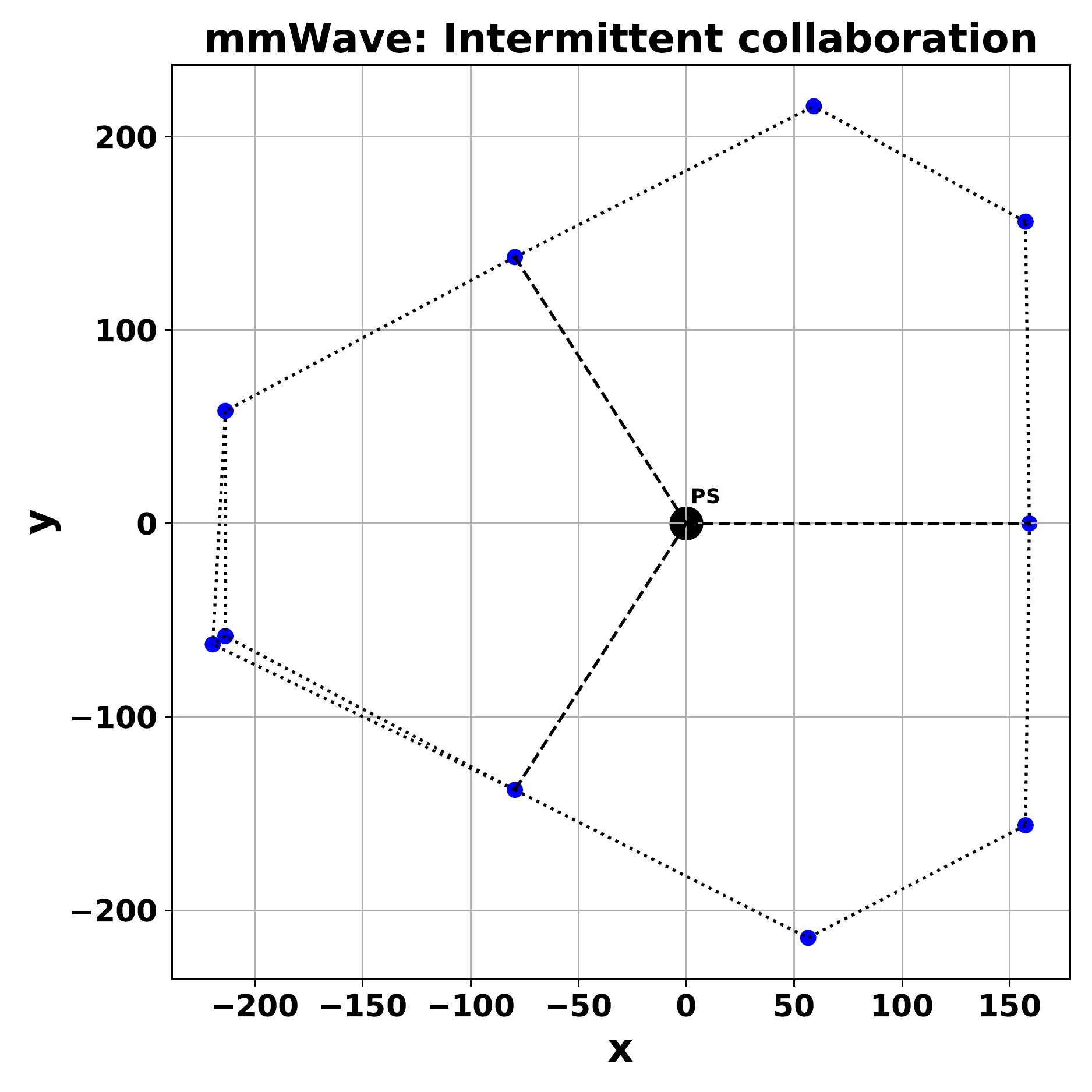}
    \caption{Intermittent client-client collaboration}
    \label{fig:mmWave_intermittent_topology}
  \end{subfigure}
  \vspace{4mm}
  \caption{Network topologies for perfect vs. intermittent client-client connectivities}
\end{figure}

\begin{figure}[t!]
\centering
  \includegraphics[width=0.75\linewidth,trim={1cm 0.8cm 1.5cm 1.8cm},clip]{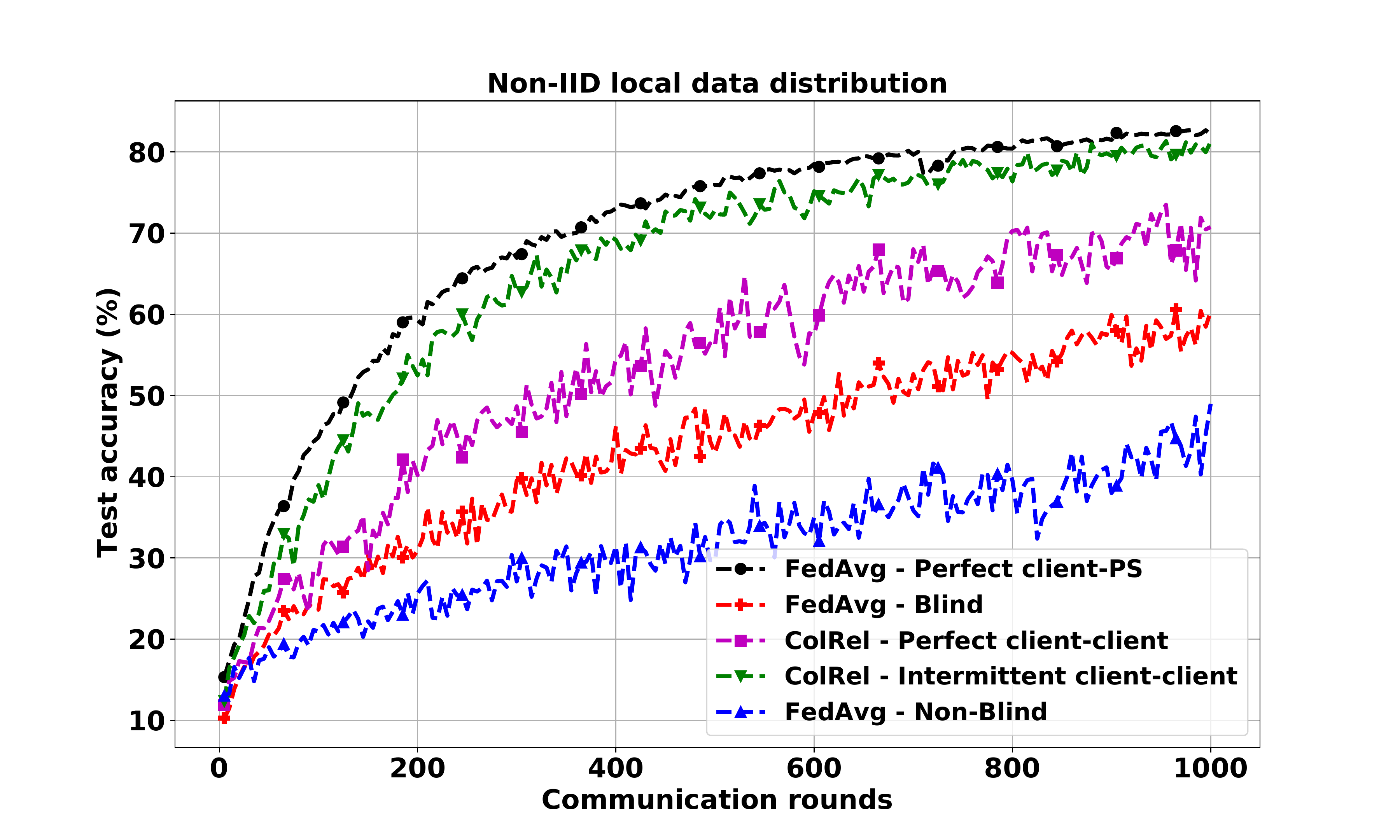}
  \caption{Improved convergence due to increased collaboration with intermittent client-client connectivity}
  \label{fig:mmWave_cluster_sortpart3_perf_int_compare_topology}
\end{figure}

In Fig. \ref{fig:mmWave_perfect_topology} we allow cooperation only among clients that are connected through permanent links. We use a threshold-based connectivity relation where $p_{ij}=1$ if $d_{ij}>d_{\text{th}}$, and $p_{ij}=0$ otherwise, where $d_{\text{th}}$ is the distance that satisfies $\mathrm{e}^{-d_{\text{th}}/30 + 5.2}=0.99$.
The PS is located at the origin and the clients are distributed in a way such that only three of them can communicate with the PS.
Similarly, in Fig. \ref{fig:mmWave_intermittent_topology}, the connectivities between clients are intermittent.
To avoid links that are too unreliable, if $p_{ij} < 0.5$, we consider those clients not to collaborate.
Note that Fig. \ref{fig:mmWave_intermittent_topology} has a few additional links compared to Fig. \ref{fig:mmWave_perfect_topology}.
In Fig. \ref{fig:mmWave_cluster_sortpart3_perf_int_compare_topology}, we observe that allowing intermittent collaboration amongst clients results in improved convergence rate of training.
Moreover, both perfect as well as intermittent connectivity outperform FedAvg without collaboration.

\section{Conclusion} \label{sec:conclusion}

FL has been proposed as an effective distributed learning strategy for mobile devices to train a common model with localized data. However, mobile devices may suffer from weak connectivity for extended durations, especially at the network edge. Our goal in this paper is to mitigate the detrimental effect of clients'  intermittent connectivity  on the training accuracy of FEEL systems.
For this purpose, we proposed a collaborative relaying strategy, called \textsc{ColRel}, which exploits the connections between clients to relay potentially missing model updates to the PS due to blocked clients. 
Our  algorithm allows the PS to receive an unbiased estimate of the model update, which would not be possible without relaying. 
We optimized the consensus weights at each client to improve the rate of convergence. 
Our proposed approach can be implemented even when the PS is blind to the identities of clients which successfully communicate with it at each round. 
Numerical results showed the improvement in training accuracy and convergence time that our approach provides under various setting, including IID and non-IID data distributions, different communication graph topologies, as well as blind and non-blind PSs.

\begin{appendices}
\section{}\label{appnd:proof:theorem_convergence_rate}
 Before proving Theorem \ref{theorem:expected_distance_to_opt}, we first introduce the following notation and lemmas.
 Denote:
 \begin{flalign}
 \overline{\xv}^{(r+1)} \triangleq \frac{1}{n}\sum_{i\in[n]}\xv_i^{(r,\mathcal{T})}.
 \end{flalign}
 \begin{lemma} \label{lemma:basic_sum_inner_prod_0}
 Under condition \eqref{eq:alpha_sum_unbiassed_contraint}, for every round $r\geq0$ we have 
\begin{flalign}
\mathbb{E}\left[\left\|\xv^{(r+1)}-x^{\star}\right\|^2\right] = \mathbb{E} \left[\left\|\xv^{(r+1)}-\overline{\xv}^{(r+1)}\right\|^2\right]+\mathbb{E} \left[\left\|\overline{\xv}^{(r+1)}-\xv^{\star}\right\|^2\right].
\end{flalign}
 \end{lemma}
 We prove Lemma \ref{lemma:basic_sum_inner_prod_0} in Appendix \ref{append_lemma:basic_sum_inner_prod_0}.
 
 \begin{lemma}\label{lemma:exp_dist_ave2optimal}
Under Assumptions \ref{assumption:Lipschitz}-\ref{assumption:positive_alpha} and condition \eqref{eq:alpha_sum_unbiassed_contraint}, for every $r\geq0$ we have 
\begin{flalign*}
&\mathbb{E} \left[\left\|\overline{\xv}^{(r+1)}-\xv^{\star}\right\|^2\right]\leq\nonumber\\ &\quad(1+n\eta_{r}^2)(1-\mu\eta_r)^{\mathcal{T}}\mathbb{E}\left[\|\xv^{(r)}-\xv^{\star}\|^2\right]
+\mathcal{T}(\mathcal{T}-1)^2L^2\frac{\sigma^2}{n}e\eta_r^2
+\mathcal{T}^2\frac{\sigma^2}{n}\eta_r^2
+\mathcal{T}^2(\mathcal{T}-1)^2L^2\sigma^2e\eta_r^4,
\end{flalign*}
for every positive $\eta_r$ such that $\eta_r\leq \min\left\{\frac{\mu}{L^2},\frac{1}{L\mathcal{T}}\right\}$.
\end{lemma}

\begin{proof}
 The proof of this lemma follows directly from Lemma 2 in \cite{reisizadeh2020fedpaq}. 
 \end{proof}

\begin{lemma}\label{lemma_max_exp_diff_i_r}
Under Assumptions \ref{assumption:Lipschitz}-\ref{assumption:positive_alpha}  and condition \eqref{eq:alpha_sum_unbiassed_contraint}, for every $r\geq0$ we have that
\begin{flalign*}
&\mathbb{E} \left[\left\|\xv_i^{(r,\mathcal{T})}-\xv^{(r)}\right\|^2\right]\leq 2{\mathcal{T}}^2L^2\eta_r^2\mathbb{E}\left[\|\xv^{(r)}-\xv^{\star}\|^2\right]
+2{\mathcal{T}}^2\sigma^2\eta_r^2
+2(\mathcal{T}-1)\mathcal{T}^2L^2\sigma^2e\eta_r^4,
\end{flalign*}
for every positive $\eta_r$ such that $\eta_r\leq \min\left\{\frac{\mu}{L^2},\frac{1}{L\mathcal{T}}\right\}$.
\end{lemma}

\begin{proof}
The proof of this lemma follows directly from (57) in the proof of  \cite[Lemma 3]{reisizadeh2020fedpaq}. 
\end{proof}

\begin{lemma}\label{lemma:dist_transmit2ave}
Under Assumptions \ref{assumption:Lipschitz}-\ref{assumption:positive_alpha}  and condition \eqref{eq:alpha_sum_unbiassed_contraint}, for every $r\geq0$ we have that
\begin{flalign*}
&\mathbb{E} \left[\left\|\xv^{(r+1)}-\overline{\xv}^{(r+1)}\right\|^2\right]\leq
\frac{2{\mathcal{T}}^2L^2\eta_r^2\mathbb{E}\left[\|\xv^{(r)}-\xv^{\star}\|^2\right]
+2{\mathcal{T}}^2\sigma^2\eta_r^2
+2(\mathcal{T}-1)\mathcal{T}^2L^2\sigma^2e\eta_r^4}{n^2}S(\boldsymbol{p},\boldsymbol{P},\boldsymbol\Alpha),
\end{flalign*}
for every positive $\eta_r$ such that $\eta_r\leq \min\left\{\frac{\mu}{L^2},\frac{1}{L\mathcal{T}}\right\}$.
\end{lemma}
 We prove Lemma \ref{lemma:dist_transmit2ave} in Appendix \ref{append_lemma:dist_transmit2ave}.

 \begin{proof}[Proof of Theorem \ref{theorem:expected_distance_to_opt}]
 By  Lemmas \ref{lemma:basic_sum_inner_prod_0}-\ref{lemma:dist_transmit2ave}, for  every positive $\eta_r$ that satisfies $\eta_r\leq \min\left\{\frac{\mu}{L^2},\frac{1}{L\mathcal{T}}\right\}$, the following holds
 \begin{flalign*}
&\mathbb{E}\left[\left\|\xv^{(r+1)}-x^{\star}\right\|^2\right] \leq (1+n\eta_{r}^2)(1-\mu\eta_r)^{\mathcal{T}}\mathbb{E}\left[\|\xv^{(r)}-\xv^{\star}\|^2\right]
+\mathcal{T}(\mathcal{T}-1)^2L^2\frac{\sigma^2}{n}e\eta_r^2
+\mathcal{T}^2\frac{\sigma^2}{n}\eta_r^2
\nonumber\\
&
+\mathcal{T}^2(\mathcal{T}-1)^2L^2\sigma^2e\eta_r^4
+\frac{2{\mathcal{T}}^2L^2\eta_r^2\mathbb{E}\left[\|\xv^{(r)}-\xv^{\star}\|^2\right]
+2{\mathcal{T}}^2\sigma^2\eta_r^2
+2(\mathcal{T}-1)\mathcal{T}^2L^2\sigma^2e\eta_r^4}{n^2}S(\boldsymbol{p},\boldsymbol{P},\boldsymbol\Alpha).
\end{flalign*}

Recall the notation $B(\boldsymbol{p},\boldsymbol{P},\boldsymbol\Alpha) =\frac{2L^2}{n^2}S(\boldsymbol{p},\boldsymbol{P},\boldsymbol\Alpha)$ and denote,
\begin{align*}
    C(\eta_{r},\boldsymbol{p},\boldsymbol{P},\boldsymbol\Alpha) &\triangleq (1+n\eta_{r}^2)(1-\mu\eta_r)^{\mathcal{T}}+B(\boldsymbol{p},\boldsymbol{P},\boldsymbol\Alpha)\mathcal{T}^2\eta_r^2,\quad
    C_1(\boldsymbol{p},\boldsymbol{P},\boldsymbol\Alpha) \triangleq \frac{4^2}{\mu^2}\cdot \frac{2\sigma^2}{n^2}S(\boldsymbol{p},\boldsymbol{P},\boldsymbol\Alpha)\nonumber\\
    C_2 &\triangleq \frac{4^2}{\mu^2}\cdot L^2\frac{\sigma^2}{n}e,\quad
    C_3(\boldsymbol{p},\boldsymbol{P},\boldsymbol\Alpha) \triangleq \frac{4^4}{\mu^4}\cdot\left( L^2\sigma^2e+\frac{2L^2\sigma^2e}{n^2}S(\boldsymbol{p},\boldsymbol{P},\boldsymbol\Alpha)\right).
\end{align*}
Therefore, 
\begin{flalign}\label{eq:dependency_on_p_alpha}
&\mathbb{E}\left[\left\|\xv^{(r+1)}-x^{\star}\right\|^2\right] \leq C(\eta_{r},\boldsymbol{p},\boldsymbol{P},\boldsymbol\Alpha)\cdot \mathbb{E}\left[\left\|\xv^{(r)}-x^{\star}\right\|^2\right]\nonumber\\
&\hspace{1.5cm}
+\left[\frac{\mu^2}{4^2}C_1(\boldsymbol{p},\boldsymbol{P},\boldsymbol\Alpha)\cdot\mathcal{T}^2
+\frac{\mu^2}{4^2} C_2\cdot\mathcal{T}(\mathcal{T}-1)^2\right]\eta_r^2+\frac{\mu^4}{4^4} C_3(\boldsymbol{p},\boldsymbol{P},\boldsymbol\Alpha)\cdot\mathcal{T}^2(\mathcal{T}-1)\eta_r^4.
\end{flalign}

We can conclude the proof using a similar approach to that presented in \cite{{reisizadeh2020fedpaq}}. 
First, we upper bound the term $C(\eta_{r},\boldsymbol{p},\boldsymbol\Alpha)$.
Since $\eta_r=\frac{4\mu^{-1}}{r\mathcal{T}+1}\leq\frac{1}{\mu}$ for every $r\geq r_0$, we can use the upper bound $\left(1-\frac{x}{m}\right)^m\leq e^{-x}$ for every $x\leq m$:
\[(1-\mu\eta_k)^{\mathcal{T}}=\left(1-\frac{\mathcal{T}\mu\eta_k}{\mathcal{T}}\right)\leq e^{-\mu \mathcal{T}\eta_r}.\]
Now, using the bound  $e^{-x}\leq1-x+x^2$ for every $x\geq 0$, we can conclude that
$
(1-\mu\eta_k)^{\mathcal{T}}\leq 1-\mu\mathcal{T}\eta_r+\mu^2\mathcal{T}^2\eta_r^2$.
It follows that 
\begin{flalign*}
C(\eta_{r},\boldsymbol{p},\boldsymbol{P},\boldsymbol\Alpha&)\leq (1+n\eta_r^2)(1-\mu\mathcal{T}\eta_r+\mu^2\mathcal{T}^2\eta_r^2)+B(\boldsymbol{p},\boldsymbol{P},\boldsymbol\Alpha)\mathcal{T}^2\eta_r^2\nonumber\\
&=n\eta_r^2(1-\mu\mathcal{T}\eta_r+\mu^2\mathcal{T}^2\eta_r^2)+1-\mu\mathcal{T}\eta_r+\mathcal{T}^2\eta_r^2\left(\mu^2+B(\boldsymbol{p},\boldsymbol{P},\boldsymbol\Alpha)\right).
\end{flalign*}

Now, since  $\eta_r=\frac{4\mu^{-1}}{r\mathcal{T}+1}\leq\frac{1}{\mu}$ and $r\geq r_0$ we have 
\begin{flalign*}
&\mu^2+B(\boldsymbol{p},\boldsymbol{P},\boldsymbol\Alpha)\leq \frac{\mu}{4\eta_r},\quad
\eta_r\leq \frac{\mu\mathcal{T}}{4n}, \quad\text{and} \quad 0\leq \mu\mathcal{T}\eta_r\leq 1.
\end{flalign*}
The maximal value of the function $1-x+x^2$ in the interval $x\in[0,1]$ is $1$, therefore, $1-\mu\mathcal{T}\eta_r+\mu^2\mathcal{T}^2\eta_r^2\leq1$.
It follows that $C(\eta_{r},\boldsymbol{p},\boldsymbol{P},\boldsymbol\Alpha)\leq 1-\frac{1}{2}\mu\mathcal{T}\eta_r$ and
\begin{flalign}\label{eq:dependency_on_p_alpha_clean}
&\mathbb{E}\left[\left\|\xv^{(r+1)}-x^{\star}\right\|^2\right] \leq \left(1-\frac{1}{2}\mu\mathcal{T}\eta_r\right)\cdot \mathbb{E}\left[\left\|\xv^{(r)}-x^{\star}\right\|^2\right]\nonumber\\
&\qquad+\left[\frac{\mu^2}{4^2}C_1(\boldsymbol{p},\boldsymbol{P},\boldsymbol\Alpha)\cdot\mathcal{T}^2
+\frac{\mu^2}{4^2} C_2\cdot\mathcal{T}(\mathcal{T}-1)^2\right]\eta_r^2+\frac{\mu^4}{4^4} C_3(\boldsymbol{p},\boldsymbol{P},\boldsymbol\Alpha)\cdot\mathcal{T}^2(\mathcal{T}-1)\eta_r^4.
\end{flalign}
Substituting $\eta_r=\frac{4\mu^{-1}}{r\mathcal{T}+1}$ we can conclude the proof by using  \cite[Lemma 5]{reisizadeh2020fedpaq} when replacing $k$ with $r$ and using the constants $k_1=\frac{1}{\mathcal{T}}$, $a=C_1(\boldsymbol{p},\boldsymbol{P},\boldsymbol\Alpha)+C_2\cdot\frac{(\mathcal{T}-1)^2}{\mathcal{T}}$ and $b=C_3\cdot\frac{(\mathcal{T}-1)}{\mathcal{T}^2}$.
\end{proof}

\section{}
\label{append_lemma:basic_sum_inner_prod_0}
\begin{proof}[Proof of Lemma \ref{lemma:basic_sum_inner_prod_0}]
Since $x\in\mathbb{R}^d$, for every $r\geq0$ we have
\begin{flalign*}
\mathbb{E}\left[\left\|\xv^{(r+1)}-x^{\star}\right\|^2\right]  &= 
\mathbb{E} \left[\left\|\xv^{(r+1)}-\overline{\xv}^{(r+1)}\right\|^2\right]+\mathbb{E} \left[\left\|\overline{\xv}^{(r+1)}-\xv^{\star}\right\|^2\right]\nonumber\\
&\quad+2\mathbb{E}\left(
\left\langle \xv^{(r+1)}-\overline{\xv}^{(r+1)},\overline{\xv}^{(r+1)}-\xv^{\star}\right\rangle\right).
\end{flalign*}

Since  $\overline{\xv}^{(r+1)}$ and $\xv^{\star}$ are deterministic functions of $(\xv_i^{(r,\mathcal{T})})_{i\in[n]}$, by the law of total expectation
\begin{flalign*}
\mathbb{E}\left(\langle \xv^{(r+1)}-\overline{\xv}^{(r+1)},\overline{\xv}^{(r+1)}-\xv^{\star}\rangle\right) 
&= \mathbb{E}\left[\left\langle \mathbb{E}\left[\xv^{(r+1)}\Big|\left(\xv_i^{(r,\mathcal{T})}\right)_{i\in[n]},\xv^{(r)}\right]-\overline{\xv}^{(r+1)},\overline{x}^{(r+1)}-\xv^{\star}\right\rangle \right].
\end{flalign*}
Now, we calculate the conditional expectation
\begin{flalign*}
&\mathbb{E}\left[\xv^{(r+1)}\Big|\left(\xv_i^{(r,\mathcal{T})}\right)_{i\in[n]},\xv^{(r)}\right]=
\mathbb{E}\left[\xv^{(r+1)}\Big|\left(\xv_i^{(r,\mathcal{T})}\right)_{i\in[n]},\xv^{(r)}\right]\nonumber\\
&\quad=
\mathbb{E}\left[\xv_i^{(r)}+\frac{1}{n}\sum_{i\in[n]}\tau_i(r+1)\sum_{j\in[n]} \tau_{j  i}(r+1)\alpha_{ij}\left(\xv_j^{(r,\mathcal{T})}-\xv^{(r)}\right)\Big|\left(\xv_i^{(r,\mathcal{T})},\right)_{i\in[n]},\xv^{(r)}\right]\nonumber\\
&\quad\stackrel{(a)}{=}
\xv^{(r)}-\frac{1}{n}\sum_{i\in[n]}p_i\sum_{j\in[n]} p_{j  i}\alpha_{ij}\xv_j^{(r)}+\frac{1}{n}\sum_{i\in[n]}p_i\sum_{j\in[n]} p_{j  i}\alpha_{ij}\xv_j^{(r,\mathcal{T})}\nonumber\\
&\quad \stackrel{(b)}{=}\frac{1}{n}\sum_{i\in[n]}\xv_i^{(r,\mathcal{T})}\triangleq \overline{\xv}^{(r+1)},
\end{flalign*}
 where $(a)$ follows since $\mathbb{E}[\tau_{i}(r+1)]=p_i$ and $\mathbb{E}[\tau_{i  j}(r+1)]=p_{i  j}$,  and $(b)$ follows from \eqref{eq:alpha_sum_unbiassed_contraint}.
\end{proof}

\section{}\label{append_lemma:dist_transmit2ave}
\begin{proof}[Proof of Lemma \ref{lemma:dist_transmit2ave}]
First, we observe that
\begin{flalign*}
&\mathbb{E} \left[\left\|\xv^{(r+1)}-\overline{\xv}^{(r+1)}\right\|^2\right] = \mathbb{E}\left[\left\|\xv^{(r)}+\frac{1}{n}\sum_{i\in[n]}\tau_i(r+1)\sum_{j\in[n]} \tau_{j  i}(r+1) \alpha_{ij}\left(\xv_j^{(r,\mathcal{T})}-\xv^{(r)}\right)-\frac{1}{n}\sum_{i \in [n]}\xv_j^{(r,\mathcal{T})}\right\|^2\right]\nonumber\\
&=\mathbb{E}\left[\left\|\xv^{(r)}-\frac{1}{n}\sum_{i\in[n]}\tau_i(r+1)\sum_{j\in[n]}\tau_{j  i}(r+1) \alpha_{ij}\xv^{(r)}\right.\right. \nonumber\\
&\hspace{3cm}\left.\left.+\frac{1}{n}\sum_{i\in[n]}\tau_i(r+1)\sum_{j\in[n]} \tau_{j  i}(r+1)\alpha_{ij}\xv_j^{(r,\mathcal{T})}-\frac{1}{n}\sum_{i \in [n]}\xv_i^{(r,\mathcal{T})}\right\|^2\right]\nonumber\\
&=\mathbb{E}\left[\left\|\xv^{(r)}-\frac{1}{n}\sum_{i,j\in[n]} \tau_j(r+1)\tau_{i  j}(r+1)\alpha_{ji}\xv^{(r)}+\frac{1}{n}\sum_{i,j\in[n]} \tau_j(r+1)\tau_{i  j}(r+1)\alpha_{ji}\xv_i^{(r,\mathcal{T})}-\frac{1}{n}\sum_{i \in [n]}\xv_i^{(r,\mathcal{T})}\right\|^2\right]\nonumber\\
&=\mathbb{E}\left[\left\|\frac{1}{n}\sum_{i\in[n]}\left(\sum_{j\in[n]} \tau_j(r+1)\tau_{i  j}(r+1)\alpha_{ji}-1\right)\xv_i^{(r,\mathcal{T})}-\frac{1}{n}\sum_{i\in[n]}\left(\sum_{j\in[n]} \tau_j(r+1)\tau_{i  j}(r+1)\alpha_{ji}-1\right)\xv^{(r)}\right\|^2\right]\nonumber\\
&=\frac{1}{n^2}\cdot\mathbb{E}\left[\left\|\sum_{i\in[n]}\left(\sum_{j\in[n]} \tau_j(r+1)\tau_{i  j}(r+1)\alpha_{ji}-1\right)\left(\xv_i^{(r,\mathcal{T})}-\xv^{(r)}\right)\right\|^2\right]\nonumber\\
&=\frac{1}{n^2}\sum_{i\in[n]}\mathbb{E}\left[\left(\sum_{j\in[n]} \tau_j(r+1)\tau_{i  j}(r+1)\alpha_{ji}-1\right)^2\right]\mathbb{E}\left[\left\|\left(\xv_i^{(r,\mathcal{T})}-\xv^{(r)}\right)\right\|^2\right]\nonumber\\
&\qquad+\frac{1}{n^2}\sum_{i,l\in[n]:i\neq l}
\mathbb{E}\left[\left(\sum_{j\in[n]} \tau_j(r+1)\tau_{i  j}(r+1)\alpha_{ji}-1\right)
\left(\sum_{m\in[n]} \tau_m(r+1)\tau_{l  m}(r+1)\alpha_{ml}-1\right)\right]\nonumber\\
&\hspace{7cm}\cdot \mathbb{E}\left[\left\langle\xv_i^{(r,\mathcal{T})}-\xv^{(r)},
\xv_l^{(r,\mathcal{T})}-\xv^{(r)}\right\rangle\right],
\end{flalign*}
where the last equality follows since the random variables $\tau_i(r+1),\:i\in[n]$ and $\tau_{i  j}(r+1),\:i,j\in[n]$ are statistically independent of  the random vectors $\xv_i^{(r,\mathcal{T})},\:i\in[n]$.
Now,
\begin{flalign*}
&\mathbb{E}\left[\left(\sum_{j\in[n]} \tau_j(r+1)\tau_{i  j}(r+1)\alpha_{ji}-1\right)^2\right]\nonumber\\
&=\sum_{j\in[n]} \mathbb{E}\left[\tau_{j}^2(r+1)\tau_{i  j}^2(r+1)\alpha^2_{ji}\right]
+\sum_{j_1,j_2\in[n]:j_1\neq j_2} \mathbb{E}\left[\tau_{j_1}(r+1)\tau_{j_2}(r+1)\tau_{i  j_1}(r+1)\tau_{i  j_2}(r+1)\alpha_{j_1i}\alpha_{j_2i}\right]\nonumber\\
&\quad-2\sum_{j\in[n]}\mathbb{E}\left[\tau_{j}(r+1)\tau_{i  j}(r+1)\alpha_{ji}\right]+1
\nonumber\\
&= \sum_{j\in[n]}p_jp_{i  j}\alpha^2_{ji}
+\sum_{j_1,j_2\in[n]:j_1\neq j_2}p_{j_1}p_{j_2}p_{i  j_1}p_{i  j_2}\alpha_{j_1i}\alpha_{j_2i}
-2\underbrace{\sum_{j\in[n]}p_{j}p_{i  j}\alpha_{ji}}_{=1}+1\nonumber\\
&=\sum_{j\in[n]}p_jp_{i  j}\alpha^2_{ji} +\underbrace{\left(\sum_{j\in[n]}p_jp_{i  j}\alpha_{ji}\right)^2}_{=1}-\sum_{j\in[n]}p^2_jp^2_{i  j}\alpha^2_{ji}-1
=\sum_{j\in[n]}p_jp_{i  j}(1-p_jp_{i  j})\alpha^2_{ji}.
\end{flalign*}

Similarly, for every $l\in[n]$ such that $l\neq i$
\begin{flalign*}
&\mathbb{E}\left[\left(\sum_{j\in[n]} \tau_j(r+1)\tau_{i  j}(r+1)\alpha_{ji}-1\right)
\left(\sum_{m\in[n]} \tau_m(r+1)\tau_{l  m}(r+1)\alpha_{ml}-1\right)\right]\nonumber\\
&\quad= \mathbb{E}\left[\sum_{j,m\in[n]} \tau_j(r+1)\tau_m(r+1)\tau_{i  j}(r+1)\tau_{l  m}(r+1)\alpha_{ji}\alpha_{ml}\right]\nonumber\\
&\qquad- \underbrace{\mathbb{E}\left[\sum_{j\in[n]} \tau_j(r+1)\tau_{i  j}(r+1)\alpha_{ji}\right]}_{=1}
-\underbrace{\mathbb{E}\left[\sum_{m\in[n]} \tau_m(r+1)\tau_{l  m}(r+1)\alpha_{ml}\right]}_{=1}+1\nonumber\\
&\quad=\mathbb{E}\left[\sum_{j\in[n]} \tau_j^2(r+1)\tau_{i  j}(r+1)\tau_{l  j}(r+1)\alpha_{ji}\alpha_{jl}\right]
+\mathbb{E}\left[\tau_l(r+1)\tau_i(r+1)\tau_{i  l}(r+1)\tau_{l  i}(r+1)\alpha_{li}\alpha_{il}\right]
\nonumber\\
&\quad+\mathbb{E}\left[\sum_{m\in[n]:m\neq l,i} \tau_l(r+1)\tau_m(r+1)\tau_{i  l}(r+1)\tau_{l  m}(r+1)\alpha_{li}\alpha_{ml}\right]
\nonumber\\
&\qquad+\mathbb{E}\left[\sum_{j\in[n]:j\neq l}\sum_{m\in[n]:m\neq j} \tau_j(r+1)\tau_m(r+1)\tau_{i  j}(r+1)\tau_{l  m}(r+1)\alpha_{ji}\alpha_{ml}\right]-1\nonumber\\
&=\sum_{j\in [n]} p_j p_{i  j}p_{l  j}\alpha_{ji}\alpha_{jl}
+p_i p_l E_{\{i,l\}}\alpha_{li}\alpha_{il}
+\sum_{m\in[n]:m\neq l,i} p_l p_m p_{i  l}p_{l  m}\alpha_{li}\alpha_{ml}+ \sum_{\substack{j\in[n]:\\ j \neq l}}\sum_{\substack{m\in[n]:\\ m \neq j}}p_j p_m p_{ij}p_{lm} \alpha_{ji}\alpha_{ml}-1
\nonumber\\
&\quad=
\left(\sum_{j\in[n]}p_jp_{i  j}\alpha_{ji}\right)\left(\sum_{m\in[n]}p_mp_{l  m}\alpha_{ml}\right)-1\nonumber\\
&\qquad -\sum_{j\in[n]}p_j^2p_{i  j}^2p_{l  j}\alpha_{ji}\alpha_{jl}+\sum_{j\in[n]}p_jp_{i  j}^2p_{l  j}\alpha_{ji}\alpha_{jl} -p_ip_lp_{i  l}p_{l  i}\alpha_{il}\alpha_{li}+p_ip_lE_{\{i,l\}}\alpha_{il}\alpha_{li}\nonumber\\
&\quad = \sum_{j\in[n]}p_j(1-p_j)p_{i  j}p_{l  j}\alpha_{ji}\alpha_{jl}+p_ip_l(E_{\{i,l\}}-p_{i  l}p_{l  i})\alpha_{il}\alpha_{li}.
\end{flalign*}

Therefore,
\begin{flalign*}
&\mathbb{E} \left[\left\|\xv^{(r+1)}-\overline{\xv}^{(r+1)}\right\|^2\right]\nonumber\\ &=\frac{1}{n^2}\sum_{i,j\in[n]}p_jp_{i  j}(1-p_jp_{i  j})\alpha^2_{ji}\mathbb{E}\left[\left\|\xv_i^{(r,\mathcal{T})}-\xv^{(r)}\right\|^2\right]\nonumber\\
&+\frac{1}{n^2}\sum_{i,l\in[n]:i\neq l}
\left( \sum_{j\in[n]}p_j(1-p_j)p_{i  j}p_{l  j}\alpha_{ji}\alpha_{jl}+p_ip_l(E_{\{i,l\}}-p_{i  l}p_{l  i})\alpha_{il}\alpha_{li}\right)
\mathbb{E}\left[\left\langle\xv_i^{(r,\mathcal{T})}-\xv^{(r)},
\xv_l^{(r,\mathcal{T})}-\xv^{(r)}\right\rangle\right]\nonumber\\
&= \frac{1}{n^2}\sum_{i,j,l\in[n]} p_j(1-p_j)p_{i  j}p_{l  j}\alpha_{ji}\alpha_{jl}\cdot
\mathbb{E}\left[\left\langle\xv_i^{(r,\mathcal{T})}-\xv^{(r)},
\xv_l^{(r,\mathcal{T})}-\xv^{(r)}\right\rangle\right]\nonumber\\
&\quad+\frac{1}{n^2}\sum_{i,j\in[n]}[p_{i  j}p_j(1-p_{i  j}p_j)-p_j(1-p_j)p^2_{i  j}]\alpha^2_{ji}\mathbb{E}\left[\left\|\left(\xv_i^{(r,\mathcal{T})}-\xv^{(r)}\right)\right\|^2\right]\nonumber\\
&\quad+\frac{1}{n^2}\sum_{i,l\in[n]:i\neq l}p_ip_l(E_{\{i,l\}}-p_{i  l}p_{l  i})\alpha_{il}\alpha_{li}\cdot
\mathbb{E}\left[\left\langle\xv_i^{(r,\mathcal{T})}-\xv^{(r)},
\xv_l^{(r,\mathcal{T})}-\xv^{(r)}\right\rangle\right]\nonumber\\
&= \frac{1}{n^2}\sum_{i,j,l\in[n]}
 p_j(1-p_j)p_{i  j}p_{l  j}\alpha_{ji}\alpha_{jl}\cdot
\mathbb{E}\left[\left\langle\xv_i^{(r,\mathcal{T})}-\xv^{(r)},
\xv_l^{(r,\mathcal{T})}-\xv^{(r)}\right\rangle\right]\nonumber\\
&\quad+\frac{1}{n^2}\sum_{i,j\in[n]}p_{i  j}p_j(1-p_{i  j})\alpha^2_{ji}\mathbb{E}\left[\left\|\left(\xv_i^{(r,\mathcal{T})}-\xv^{(r)}\right)\right\|^2\right]\nonumber\\
&\quad+\frac{1}{n^2}\sum_{i,l\in[n]}p_ip_l(E_{\{i,l\}}-p_{i  l}p_{l  i})\alpha_{il}\alpha_{li}\cdot
\mathbb{E}\left[\left\langle\xv_i^{(r,\mathcal{T})}-\xv^{(r)},
\xv_l^{(r,\mathcal{T})}-\xv^{(r)}\right\rangle\right],
\end{flalign*}
where the last equality follows since $E_{\{i,i\}}-p_{i  i}p_{i  i}=1-1\cdot 1=0$.

By Lemma \ref{lemma_max_exp_diff_i_r} and the Cauchy–Schwarz inequality we have for each $i\in[n]$ 
\begin{flalign*}
\left|\mathbb{E}\left[\left\langle\xv_i^{(r,\mathcal{T})}-\xv^{(r)},
\xv_l^{(r,\mathcal{T})}-\xv^{(r)}\right\rangle\right]\right|
&\quad\leq \sqrt{\mathbb{E}\left[\left\|\xv_i^{(r,\mathcal{T})}-\xv^{(r)}\right\|^2\right]}\cdot\sqrt{\mathbb{E}\left[\left\|\xv_l^{(r,\mathcal{T})}-\xv^{(r)}\right\|^2\right]}\nonumber\\
&\quad\leq 2{\mathcal{T}}^2L^2\eta_r^2E\|\xv^{(r)}-\xv^{\star}\|^2
+2{\mathcal{T}}^2\sigma^2\eta_r^2
+2(\mathcal{T}-1)\mathcal{T}^2L^2\sigma^2e\eta_r^4.
\end{flalign*}
Consequently, it follows from Assumption \ref{assumption:positive_alpha} and the definition of $S(\boldsymbol{p},\boldsymbol{P},\boldsymbol\Alpha)$ that
\begin{flalign*}
&\mathbb{E}\left[ \left\|\xv^{(r+1)}-\overline{\xv}^{(r+1)}\right\|^2\right]\leq
\frac{2{\mathcal{T}}^2L^2\eta_r^2\mathbb{E}\left[\|\xv^{(r)}-\xv^{\star}\|^2\right]
+2{\mathcal{T}}^2\sigma^2\eta_r^2
+2(\mathcal{T}-1)\mathcal{T}^2L^2\sigma^2e\eta_r^4}{n^2}S(\boldsymbol{p},\boldsymbol{P},\boldsymbol\Alpha).
\end{flalign*}
\end{proof}
\vspace{-1cm}

\section{}\label{append:proof_lemma_conex_S_func_A}
To prove Lemma \ref{lemma:conex_S_func_A} we first present the following auxiliary result.
\begin{lemma}\label{lemma:conex_square}
Let $\yv,\cv\in\mathbb{R}^{\tilde{d}}$ where $\tilde{d}\in\mathbbm{N}_+$,
and  $h_{\cv}(\yv)\triangleq\left(\sum_{i=1}^{\tilde{d}} \cv_i\yv_i\right)^2$. Function $h_{\cv}(\yv)$ is convex.
\end{lemma}

\begin{proof}[Proof of Lemma \ref{lemma:conex_square}]
We prove this lemma by using the definition of a convex function, namely showing that 
$h_{\cv}(\lambda \yv+(1-\lambda) \zv)\leq \lambda h_{\cv}(\yv)+(1-\lambda)h_{\cv}(\zv)$,
for every $\yv,\zv,\cv\in\mathbb{R}^{\tilde{d}}$ and $\lambda\in[0,1]$:
\begin{flalign*}
&\lambda h_{\cv}(\yv)+(1-\lambda)h_{\cv}(\zv)-h_{\cv}(\lambda \yv+(1-\lambda) \zv)\nonumber\\
&=\lambda\left(\sum_{i=1}^{\tilde{d}} \cv_i\yv_i\right)^2+(1-\lambda)\left(\sum_{i=1}^{\tilde{d}} \cv_i\zv_i\right)^2
-\lambda^2\left(\sum_{i=1}^{\tilde{d}} \cv_i\yv_i\right)^2\nonumber\\
&\hspace{1cm}-2\lambda(1-\lambda)\left(\sum_{i=1}^{\tilde{d}} \cv_i\yv_i\right)\left(\sum_{i=1}^{\tilde{d}} \cv_i\zv_i\right)-(1-\lambda)^2\left(\sum_{i=1}^{\tilde{d}} \cv_i\zv_i\right)^2\nonumber\\
&=\lambda(1-\lambda)\left(\sum_{i=1}^{\tilde{d}} \cv_i\yv_i\right)^2+\lambda(1-\lambda)\left(\sum_{i=1}^{\tilde{d}} \cv_i\zv_i\right)^2-2\lambda(1-\lambda)\left(\sum_{i=1}^{\tilde{d}} \cv_i\yv_i\right)\left(\sum_{i=1}^{\tilde{d}} \cv_i\zv_i\right)\nonumber\\
&=\lambda(1-\lambda)\left[\sum_{i=1}^{\tilde{d}} \cv_i\yv_i-\sum_{i=1}^{\tilde{d}} \cv_i\zv_i\right]^2\geq0,
\end{flalign*}
where the last inequality follows since $\yv,\zv\in\mathbb{R}^{\tilde{d}}$ and $\lambda\in[0,1]$.
\end{proof}

\begin{proof}[Proof of Lemma \ref{lemma:conex_S_func_A}]
First,  note that $S(\boldsymbol{p},\boldsymbol{P},\boldsymbol\Alpha)\leq\overline{S}(\boldsymbol{p},\boldsymbol{P},\boldsymbol\Alpha)$ since $xy\leq \frac{1}{2}(x^2+y^2),\:\forall\:x,y\in\mathbb{R}$.

We prove that $\overline{S}(\boldsymbol{p},\boldsymbol{P},\boldsymbol\Alpha)$ is convex with respect to $\boldsymbol\Alpha$ by rewriting its first summation as a nonnegative weighted sum of convex quadratic functions: 
\begin{flalign*}
\sum_{i,l\in[n]}\sum_{j\in[n]} p_j(1-p_j)p_{i  j}p_{l  j}\alpha_{ji}\alpha_{jl}
&=\sum_{j\in[n]}p_j(1-p_j)\sum_{i,l\in[n]}p_{i  j}p_{l  j}\alpha_{ji}\alpha_{jl}=\sum_{j\in[n]}p_j(1-p_j)\left(\sum_{i\in[n]}p_{i  j}\alpha_{ji}\right)^2,
\end{flalign*}
where the equality follows since if $j\in\mathcal{N}_i\cup\{i\}$ and $j\in\mathcal{N}_{l}\cup\{l\}$, then  $j\in\mathcal{N}_{il}$.
Now, since the function $\left(\sum_{i\in[n]}p_{i  j}\alpha_{ji}\right)^2$ is convex in $\boldsymbol\Alpha$ for every $\boldsymbol{P}$ (see Lemma \ref{lemma:conex_square}), and since $p_j,p_{i  j}\in[0,1]$ for every $i,j\in[n]$, and \my{$E_{\{i,j\}}\geq p_{ij}p_{ji}$ for every $i,j\in[n]$,}
the function $\overline{S}(\boldsymbol{p},\boldsymbol{P}, \boldsymbol\Alpha)$ is convex.
\end{proof}

\section{}\label{append:Solve_alpha_allocation_lagrange}
First, observe that  we can set
$\alpha_{ji}=0$ for every $j$ such that $p_{i  j}p_j=0$.
Additionally, if $\max_{k\in[n]}\{p_kp_{i  k}\}=1$, then we can set $\alpha_{ji}=\mathbbm{1}_{\{p_{i  j}p_j=1\}}\cdot(\sum_{k\in[n]}\mathbbm{1}_{\{p_kp_{i  k}=1\}})^{-1}$.
Therefore, hereafter we assume that $i$ is such that $\max_{k\in[n]}\{p_kp_{i  k}\}<1$ and that $j$ is such that $p_{i  j}p_j\in(0,1)$.
We proceed to solve each of the problems \eqref{eq:opt_alpha_allocation_iterative_upper} and \eqref{eq:opt_alpha_allocation_iterative}, respectively.

\subsection{Solving the Convex Optimization Problem  \eqref{eq:opt_alpha_allocation_iterative_upper}}
The Lagrangian of \eqref{eq:opt_alpha_allocation_iterative_upper} is
\begin{flalign*}
\overline{L}(\boldsymbol{\Alpha}_{i}^{(\ell)},\lambda_i) &= \sum_{j\in[n]}p_jp_{i  j}\left(1-p_jp_{i  j}\right)\alpha_{ji}^2+2\sum_{l\in[n],l\neq i}\sum_{j\in[n]} p_j(1-p_j)p_{i  j}p_{l  j}\alpha_{ji}\alpha_{jl}^{(\ell-1)}\nonumber\\
&+\sum_{j\in[n]}p_ip_j(E_{\{i,j\}}-p_{i  j}p_{j  i})\alpha^2_{ji}-\lambda_i\left(\sum_{j\in[n]} p_jp_{i  j}\alpha_{ji}-1\right)-\mu_{ji}(\alpha_{ji}).
\end{flalign*}
Additionally, 
\begin{flalign*}
\frac{\partial \overline{L}(\boldsymbol\Alpha_{i}^{(\ell)},\lambda_i)}{\partial \alpha_{ji}}&=
2p_j[p_{i  j}\left(1-p_jp_{i  j}\right)+p_i(E_{\{i,j\}}-p_{i  j}p_{j  i})]\alpha_{ji}\nonumber\\
&\quad+2p_j(1-p_j)p_{i  j}\sum_{l\in [n]:l\neq i} p_{l  j}\alpha_{jl}^{(\ell-1)}-\lambda_ip_{i  j} p_j+\mu_{ji},\\ 
\frac{\partial \overline{L}(\boldsymbol\Alpha_{i}^{(\ell)},\lambda_i)}{\partial \lambda_i} &= 1-\sum_{j\in[n]} p_jp_{i  j}\alpha_{ji},\qquad 
\frac{\partial \overline{L}(\boldsymbol\Alpha_{i}^{(\ell)},\lambda_i)}{\partial \mu_{ji}} = -\alpha_{ji}.
\end{flalign*}

It follows from the Karush–Kuhn–Tucker conditions that
\begin{flalign*}
\alpha_{ji}(\lambda_i)
&=\left(\frac{-2(1-p_j)\sum_{l\in [n]:l\neq i} p_{l  j}\alpha_{jl}^{(\ell-1)}+\lambda_i }{2[\left(1-p_jp_{i  j}\right)+p_i(E_{\{i,j\}}/p_{i  j}-p_{j  i})]}\right)^+,
\end{flalign*}
and 
$\lambda_i\geq0$ is set such that $\sum_{j\in[n]}p_jp_{i  j}\alpha_{ji}(\lambda_i)=1$.

\subsection{Solving the Convex Optimization Problem \eqref{eq:opt_alpha_allocation_iterative}}
The Lagrangian of \eqref{eq:opt_alpha_allocation_iterative} is
\begin{flalign*}
L(\boldsymbol{\Alpha}_{i}^{(\ell)},\lambda_i) &= \sum_{j\in[n]}p_jp_{i  j}\left(1-p_jp_{i  j}\right)\alpha_{ji}^2+2\sum_{l\in[n],l\neq i}\sum_{j\in[n]} p_j(1-p_j)p_{i  j}p_{l  j}\alpha_{ji}\alpha_{jl}^{(\ell-1)}\nonumber\\
&+2\sum_{j\in[n]}p_ip_j(E_{\{i,j\}}-p_{i  j}p_{j  i})\alpha_{ji}\alpha_{ij}^{(\ell-1)}-\lambda_i\left(\sum_{j\in[n]} p_jp_{i  j}\alpha_{ji}-1\right)-\mu_{ji}(\alpha_{ji}).
\end{flalign*}
Additionally, 
\begin{flalign*}
\frac{\partial L(\boldsymbol\Alpha_{i}^{(\ell)},\lambda_i)}{\partial \alpha_{ji}}&=
2p_jp_{i  j}\left(1-p_jp_{i  j}\right)+2p_j(1-p_j)p_{i  j}\sum_{l\in [n]:l\neq i} p_{l  j}\alpha_{jl}^{(\ell-1)}\nonumber\\
&\quad+2p_ip_j(E_{\{i,j\}}-p_{i  j}p_{j  i})\alpha_{ij}^{(\ell-1)}-\lambda_ip_{i  j} p_j+\mu_{ji},\\ 
\frac{\partial L(\boldsymbol\Alpha_{i}^{(\ell)},\lambda_i)}{\partial \lambda_i} &= 1-\sum_{j\in[n]} p_jp_{i  j}\alpha_{ji},\qquad 
\frac{\partial L(\boldsymbol\Alpha_{i}^{(\ell)},\lambda_i)}{\partial \mu_{ji}} = -\alpha_{ji}.
\end{flalign*}

It follows from the Karush–Kuhn–Tucker conditions that
\begin{flalign*}
\alpha_{ji}(\lambda_i)
&=\left(\frac{-2(1-p_j)\sum_{l\in [n]:l\neq i} p_{l  j}\alpha_{jl}^{(\ell-1)}-2p_i(E_{\{i,j\}}/p_{i  j}-p_{j  i})\alpha_{ij}^{(\ell-1)}+\lambda_i }{2\left(1-p_jp_{i  j}\right)}\right)^+,
\end{flalign*}
and 
$\lambda_i\geq0$ is set such that $\sum_{j\in[n]}p_jp_{i  j}\alpha_{ji}(\lambda_i)=1$.

\end{appendices}

\let\oldbibliography\thebibliography
\renewcommand{\thebibliography}[1]{\oldbibliography{#1}
\setlength{\itemsep}{0pt}} 




\end{document}